# Chemical analysis and aqueous solution properties of Charged Amphiphilic Block Copolymers PBA-b-PAA synthesized by MADIX®

M. Jacquin[a], P. Muller[a,d], R. Talingting-Pabalan[b], H. Cottet[c], J.F. Berret[a] , T. Futterer[b], O. Théodoly[a,e *]

[a]Complex Fluids Laboratory, CNRS UMR 166, 350 George Patterson Blvd, Bristol PA 19007, USA.
[b]Rhodia CRTB, 350 George Patterson Blvd, Bristol PA 19007, USA.
[c]Equipe Dynamique des Systèmes Biomoléculaires Complexes, CNRS UMR 5073, Université de Montpellier 2, place Eugène Bataillon CC 017, 34095 Montpellier Cedex 5, France.
[d]Institut Charles Sadron, 6 rue Boussingault, 67083 Strasbourg Cedex, France.
[e]Laboratoire Adhésion et Inflammation, INSERM U600, CNRS UMR 6212, Case 937, 163 Avenue de Luminy, Marseille F-13009, France ; Aix-Marseille Université, Faculté des Sciences/de Médecine ou de Pharmacie, Marseille, F-13000, France.

*corresponding author. Tel.: +33 (0)4 91 82 75 79; Fax: Tel.: +33 (0)4 91 82 75 80.
email:theodoly@marseille.inserm.fr

keywords: amphiphilic diblock copolymers, self assembling, interfacial tension

ABSTRACT: We have linked the structural and dynamic properties in aqueous solution of amphiphilic charged diblock copolymers poly(butyl acrylate)-b-poly(acrylic acid), PBA-b-PAA, synthesized by controlled radical polymerization, with the physico-chemical characteristics of the samples. Despite product imperfections, the samples self-assemble in melt and aqueous solutions as predicted by monodisperse microphase separation theory. However, the PBA core are abnormally large; the swelling of PBA cores is not due to AA (the Flory parameter $\chi_{PBA/PAA}$, determined at 0.25, means strong segregation), but to h-PBA homopolymers (content determined by Liquid Chromatography at the Point of Exclusion and Adsorption Transition LC-PEAT). Beside the dominant population of micelles detected by scattering experiments, capillary electrophoresis CE analysis permitted detection of two other populations, one of h-PAA, and the other of free PBA-b-PAA chains, that have very short PBA blocks and never self-assemble. Despite the presence of these free unimers, the self-assembly in solution was found out of equilibrium: the aggregation state is history dependant and no unimer exchange between micelles occurs over months (time-evolution SANS). The high PBA/water interfacial tension, measured at 20 mN/m, prohibits unimer exchange between micelles. PBA-*b*-PAA solution systems are neither at thermal equilibrium nor completely frozen systems: internal fractionation of individual aggregates can occur.

Keywords: diblock copolymer, micelle, interfacial tension, exchange kinetics, self assembly, polyelectrolyte, fractionation.

## 1. Introduction

Water-soluble surface active block copolymers present a great potential as alternatives or additives to common small-molecule surfactants. Typical applications of amphiphilic block copolymers enclose all small-molecule surfactant-based formulations for detergency, suspension stabilization, emulsion polymerization, wetting modification, etc. Advantages of using block copolymers rather than small-molecule surfactants are, for instance, a longer range of repulsion (steric and/or electrostatic) between surfaces covered by amphiphilic molecules or a better anchoring of the stabilizers on the surfaces. However, the potential of amphiphilic diblocks for applications has long been hindered by the difficulty to synthesize block copolymers in a simple and cheap enough way for industrial-scale production. Anionic polymerization is the best controlled way to produce block copolymers, with index of polydispersities as low as 1.01. Such samples have proven very precious for academic research but their synthesis conditions are too expensive for most commercial applications. Pluronics® polymers, produced industrially from an anionic process, are an exception to this rule. Still, side reactions with PPO lead to poorly controlled products and the advantage of using an anionic polymerization process is lost. The other major limitation of anionic synthesis process comes from the narrow range of chemical compounds that can be properly polymerized (PB, PS, PI, PtBA, PMMA and PEO). Fortunately, in the last decade, controlled radical polymerizations [1] like RAFT [2] and ATRP [3,4] have been developed as an alternative route to block copolymer synthesis, opening a new era for block copolymer science and industry. These processes allow relatively cheap and large-scale production of block copolymers out of a wide variety of chemistries. The cons are that (i) chains are more polydisperse and (ii) side-products are generated in larger quantities. The index of polydispersity is typically around 1.2-2, which is much better than the polydispersity of radical polymerization products (typically 3), but significantly higher than the one of anionic polymerization products (as low as 1.01 in best cases). Side products are mainly homopolymers of each block and, to a lesser extent, multiblocks. As a consequence, physico-chemical properties of these products are expected to be strongly dependent on the composition of the products and are expected to differ from the targeted properties of a pure block copolymer sample. This subject has been largely overlooked in the literature and systematically discussed in this paper. We investigate here the physical properties in melt and in aqueous solution of diblock copolymers produced by controlled radical polymerization. Our aim is to understand the non ideal physical properties on the basis of a precise chemical characterization of the products. This approach appeared sane and necessary for controlled radical polymerization samples, which are known to be imperfect. However, the same approach could also be enlightening with anionic polymerization samples, whose imperfections are not always negligible [5-11]. We have investigated samples of poly(butyl acrylate)-b-poly(acrylic acid), or PBA-b-PAA, synthesized by the Rhodia patented MADIX® process [12]. These polymers were designed for their potential surface active properties in aqueous solution, which are most valued for applications. These products have already been successfully tested in formulations to increase stability of direct and inverse emulsions. Our final goal with these PBA-b-PAA samples is to achieve an academic understanding of their surface properties, e.g. to link their properties in solution and their properties at surfaces or in the presence of small-molecule surfactant. The present paper deals with the first step of this study, i.e. the physical properties of PBA-b-PAA samples in bulk aqueous solutions. The originality of the present study is that it links the synthesis mechanism, the chemical composition of the samples (determined by GPC, NMR, Capillary Electrophoresis CE, and Liquid Chromatography at the Point of Exclusion Adsorption Transition LCPEAT), and the physical properties of the solutions (characterized by DLS, SANS, and SAXS). We will focus on the structural parameters of self-assembled aggregates (in melt and in solution), the values of CMC's in solution, and the reversibility of the self-assembly in solution. In the end, this paper gives a new insight on the self assembling properties of diblock copolymer in aqueous solution in relation to physical parameters, e.g. the interfacial tension between the PBA and water, and to chemical parameters, e.g. the presence of homopolymers.

## 2. Experimental Section

### 2.1. Synthesis.

**Materials.** Monomers butyl acrylate (Aldrich, 234923, 99%), acrylic acid (Aldrich, 147230, 99%) and deuterated butyl $d_9$-acrylate (Polymer Source, D9nBUA), initiator 2,2'-azobis(2-methylbutanenitrile) (AMBN, DuPont Vazo 67), controlling agent (2-mercaptopropionic acid, methyl ester, o-ethyl dithiocarbonate) (Rhodixan A1, Rhodia, 99% NMR purity) and solvent ethanol (Aldrich, 459844) were used without further purification. Water was purified using a Milli-Q plus water purification system.

**Process.** All polymers were synthesized via controlled radical polymerization process MADIX®. For the first block, a mixture of monomers in ethanol at a concentration of 40 wt% was bubbled with ultrapure $N_2$ and heated at 70ºC in presence of xanthate. A shot of initiator was then added to start the reaction. The ratio of the monomer mass $M_m$ introduced versus the number of Xanthate moles $n_{Xa}$ fixed the targeted molar mass $M_t$ of the polymers chains to

$$M_t = \frac{M_m}{n_{Xa}} \quad \text{Eq. [1]}$$

After completion of the first block growth, an aliquot was taken for analysis and the second monomer was added as a 40 wt% solution in ethanol. A new shot of initiator was added to start the second block. Finally, the reaction products were dialyzed versus pure water to remove ethanol and unreacted monomers. They were then freeze-dried and stored. Our samples names refer to the targeted molecular weight $M_t$ in Dalton. For instance, a PBA-b-PAA 4k-20k stands for a sample whose PBA block was targeted at 4k and PAA block at 20k.

**Optimization.** The reaction time in a realistic industrial process is overestimated to insure a complete conversion of monomers, which increases the probability of trans-esterification and chain transfer to the solvent events [13]. Monomers conversions were controlled during reaction by way of GC-Head Space and dry extract measurements. The reaction times were optimized to stop the reaction as soon as the residual amount of monomers was lower than 1% of the initial concentration (typically less than 8 hours reaction time per block). The amount of initiator versus xanthate has to be minimized to limit the production of dead chains at the end of each block synthesis. However, lowering this ratio slows down the polymerization kinetics, which increases the probability of trans-esterification by solvent events, transfer to solvent events and termination reactions. This tends to increase respectively the amount of ethylacrylate group in the samples, the amount of dead homopolymer chains and the amount of unreacted monomers. An optimum was found for a molar ratio xanthate over initiator $\frac{n_{Xanthate}}{n_{initiator}}$ of 10. In such conditions, chain terminations are negligible vs chain transfer to xanthate and the effect of trans(esterification) and transfer to solvent events are quantified in the following.

## 2.2. Chemical analysis.

**NMR**. $^{13}C$ NMR spectra were measured on polymer solutions in DMSO using a 400 MHz Inova unity from Varian. Chromium acetyl acetonate was added to the solution to decrease the relaxation time of carbons and allow quantitative measurements. Attribution of peaks depending on displacements is reported on Figure 1. Chemical shifts for all the carbonyl groups (acrylic acid (AA), ethyl acrylate (EA), and butyl acrylate (BA)) are between 170-180 ppm. The presence of EA results from trans-esterification with the solvent, i.e. reaction of BA and AA monomers with ethanol. The molar fraction of EA is equal to the ratio of the integral of the peak at 60 ppm (carbon 2) versus the integral of all carbonyl groups (carbon 1). The molar fraction of BA is equal to the ratio of the integral of peak at 64 ppm (carbon 4) versus the integral of all carbonyl groups. On

Table 1, the mass ratio of PBA versus PAA determined from NMR $X_{BA/AA}^{NMR}$ and the number fraction of ethyl acrylate groups in the sample $r_{EA}^{NMR}$ are reported for all samples.

**THF GPC**. The molecular weight distributions (cf. Figure 2) have been determined by Gel Permeation Chromatography on a series of 3 columns PLgel Mixed Beads (Polymer Laboratories) using THF as a mobile phase and refractive index as detection. Columns have been calibrated with PS standards (EasiCal PS-2, Polymer Laboratories). For block copolymers started by the PBA block, the mass distribution of the first block PBA was measured on aliquots taken at the end of the first block reaction. The mass distribution of the whole PBA-b-PAA diblock was measured on samples modified by methylation to avoid specific interactions of PAA with the column. Methylation was performed by first dissolving the diblock sample in DMF in presence of iodomethane in a 2/1 molar ratio relative to acrylic acid groups. One equivalent of tetrabutyl ammonium hydroxide relative to acrylic acid was then slowly added from a methanol solution over a period of 4 hours. The reaction was run in the dark for 48 hours at room temperature. Reaction yield, determined by $^{13}C$ NMR in $CDCl_3$, showed that more than 90% of acrylic acid groups had been esterified.

**Aqueous GPC**. Analytical measurements were done on a Waters Alliance 2690 GPC unit, with a refractive index and a light scattering detector in line (Wyatt Technology, OPTILAB and DAWN DSP). 100 μL of sample solutions in the charged state (pH around 9) were injected at a concentration of 1 wt%. The column was a Tosoh TSK-gel G5000PW 300 x 7.5 mm - 10 μm. The mobile phase was a $NaNO_3$ solution at 0.1 M, without any organic solvent, and the flow rate was fixed at 1.5 mL/min. Analytical aqueous GPC was used on the aliquots of PAA for block copolymers started by the PAA block. It was shown, by measurements performed on standards that this technique leads to an overestimation of the molecular weight of the samples and polydispersities of 2.5 for monodisperse samples. Analyses of PAA have therefore to be taken cautiously. Aqueous GPC was also performed to investigate the composition of aqueous solutions of PBA-b-PAA. These experiments were performed at the analytical and the preparative scale. Preparative GPC was performed on a Varian PREPSTAR with UV detection. The column used was TSK-Gel G 0001-05K 600 x 21.5 mm - 22 µm. The same mobile phase was used with a flow rate fixed at 8 mL/min. Our aim was to separate aggregates of chains from eventual lone chains present in aqueous solution. Working with a mobile phase exempt of organic solvent was required to avoid any perturbation of the aggregation pattern of the solutions of interest. Nevertheless, irreversible adsorption of the polymer on the columns was observed. This was a problem for a rigorous quantitative analysis of the aggregates population but not for the preparative separation.

**LC-PEAT.** Liquid Chromatography at the Point of Exclusion and Adsorption Transition is an analytical technique optimized to reach critical conditions for a given polymer, i.e. conditions where the elution time for this polymer does not depend on its molecular weight. Critical conditions have been determined for PBA [14] which allows separating homopolymer h-PBA from diblock PBA-b-PAA and homopolymer h-PAA (Figure 3). Critical conditions were achieved for PBA on a Zorbax SB C18 300 4.6x250 mm column at room temperature. The eluant was a THF/$H_2O$ mixture of volume ratio comprised between 90/10 and 89/11. Typically, 10 µL of a 1wt% solution were injected in a 1mL/min flow rate. Polymer detection was achieved using an ELSD (Evaporative Light Scattering Detector) SEDERE PL EMD-960 regulated at 40˚C, with a gas flow of 3.5-5ml/min. The detector response was calibrated for the peak of *h*-PBA, i.e. peak (1) on figure 3, by injections of known amounts of *h*-PBA 3K in the mobile phase at different known concentrations. It was checked that the calibration does not depend on the $M_w$ of the h-PBA used. The integral of the peak of h-PBA could then be used to determine the mass fraction of homopolymer h-PBA in the samples $X_{h-PBA}^{LC-PEAT}$ .

**Capillary Electrophoresis**. Capillary electrophoresis is an analytical separation technique based on the differential migration of ionic species. We used an Agilent CE apparatus to perform capillary zone electrophoresis (CZE). A high voltage, typically 16 kV, was applied to a fused silica capillary filled with a 160 mM sodium borate electrolyte, at a controlled temperature of 25°C. Capillary dimensions are 33.5 cm x 50 µm (25 cm to the detector). Injections of samples at 1 wt% were performed hydrodynamically using pressures between 17 and 40 mbar for 3 seconds. The apparent electrophoretic mobility $\mu_{app}$ is defined according to:

$$\mu_{app} = \frac{v_{app}}{E} = \frac{Ll}{Vt_{app}}$$ **Eq; [2]**

where $v_{app}$ is the apparent electrophoresis velocity, E is the electric field, L is the total capillary length, l is the effective capillary length, V is the applied voltage, and $t_{app}$ is the apparent migration time of the solute. The apparent mobility differs from the effective mobility of the solute due to the electroosmotic flow. The effective mobility is therefore defined in Eq. [3]:

$$\mu_{ep} = \mu_{app} - \mu_{eo} = \frac{Ll}{Vt_{app}} - \frac{Ll}{Vt_{eo}}$$ **Eq. [3]**

where $t_{eo}$ is the migration time of neutral molecules. UV detection was used at wavelengths of 200 nm and 290 nm, the former being sensitive to acrylate groups and the latter being specific to xanthate. Figure 4 shows that CZE allows to separate three peaks. Peak 1 corresponds to homopolymer h-PAA. This has been checked by addition of h-PAA in the samples. By default, peaks 2 and 3 correspond to all species in the sample at the exception of h-PAA. We could then deduce the mass fraction of homopolymer h-PAA in the samples $X^{CE}_{h-PAA}$. For identification of peak 2 and 3, networks of dextran have been used to separate the species vs their size, and surfactant Brij 35 has been added at a concentration of 5 mM (7-15 CMC) to detect the possible affinities of the species with surfactant (cf. section 3.3.3).

**Sample composition.** The chemical analysis described above allowed a thorough characterization of the samples. The final product is a mixture of the diblock and the two homopolymers h-PBA and h-PAA. The mass fraction of h-PBA homopolymer is given by the LC-PEAT analysis and is directly equal to $X^{LC-PEAT}_{h-PBA}$. The mass fraction of h-PAA homopolymer is obtained by capillary electrophoresis and is equal to $X^{CE}_{h-PAA}$. The mass fraction of diblock is then simply 1- $X^{LC-PEAT}_{h-PBA}$ - $X^{CE}_{h-PAA}$. The mass ratio $X_{BA/AA}$ of the diblock component of the mixture can be computed with Eq. [4]. (cf. values on Table 1):

$$X_{BA/AA}=\frac{X^{NMR}_{BA/AA}-X^{LC-PEAT}_{h-PBA}\left(1+X^{NMR}_{BA/AA}\right)}{1-X^{CE}_{h-PAA}\left(1+X^{NMR}_{BA/AA}\right)} \qquad \textbf{Eq. [4]}$$

**2.3. Physical analysis.**

**Density.** Density of h-PBA was determined by measuring the weight and the height of a h-PBA liquid column in a capillary. The capillary size was calibrated using pure water.

**Interfacial tension.** A pendant drop Rame-Hart Goniometer was used to measure interfacial tensions. Air-water interfacial tensions were measured on hanging bubbles, whose volume was kept constant over time by a step-motor activated syringe controlled by a feedback loop. The interfacial tension between h-PBA and water was measured on sessile drops of h-PBA immersed in water. All glassware has been cleaned in KOH saturated ethanol solutions for 10 minutes then rinsed thoroughly. Milli-Q water was used for all measurements.

**Light scattering measurements.** Dynamic light scattering experiments (DLS) were performed on a Brookhaven spectrometer (BI-9000AT autocorrelator), that allows acquisition over eight decades of relaxation rates from 0.1 μs to 10 s and multi-angle scans between 20° and 155°. All data were analyzed using multi-angle CONTIN method by Provencher [15]. The measurements were made at different concentrations and the final hydrodynamic radius $R_h$ values were the values obtained by extrapolation at a zero concentration (in order to avoid the influence of the second virial coefficients). All solutions were filtered on a 0.2 μm filter (Millex, Millipore) before measurements. DLS was used to study the hydrodynamic radii of polymer micelles versus ionization of the corona, i.e. pH, and ionic strength of the solution, i.e. NaCl background concentration (cf. Figure 11).

**SAXS and SANS.** Small Angle X-ray Scattering (SAXS) measurements have been performed on X21 at NSLS, Brookhaven National Laboratory, USA, and Small Angle Neutron Scattering (SANS) measurements on NG3 at NIST, Gaithersburg, USA and on PACE at LLB, Saclay, France. SANS and SAXS data have been used to determine physical characteristics of structures in melt and in aqueous solutions. For spheres in melt, form factor of polydisperse sphere coupled with a Percus-Yevick [16] structure factor have been applied to fit the data. For cylinders and lamella, we used the positions of structure peaks to identify the topology and the correlation lengths. For aqueous dispersions of spherical micelles, which include almost all the solutions considered in this work, different models have been applied. In the case of low pH (around 2), where the PAA is mostly neutral, the Daoud Cotton model [17] was applied to fit the data. It is based on a dense core and a corona whose radial concentration decreases with a power law. The parameters for the fit are the density of the core $\rho_c$, the radius of the core $R_c$, the density of the corona next to the core $\rho_1$ and the exponent of the power law n. For a neutral chain in good solvent conditions, the theoretical exponent is –4/3. This mean field approach is adapted to low angle scattering data. Alternatively, polydisperse core-shell Pedersen model adjustments [18] were used to fit the data, which is more accurate at high scattering angle because the cross-terms between each chain of the corona with the core and the other chains of the corona are taken into account. The Pedersen model considers a core of pure PBA of size $R_c$ surrounded by $N_{agg}$ Gaussian chains of PAA of molar masses $M_W$(PAA). In the case of high pH (around 9), the corona chains are highly charged and stretched. A Gaussian configuration is not relevant any more and the Pedersen model is not adapted. We have therefore applied Urchin-like models [19,20]. Urchin-like model is a combination of the mean-field term for a core (radius $R_c$) and a stretched corona (corona profile decreases with a power law of –2) and of the term resulting from $N_{agg}$ stretched chains of length $L_{rod}$. For our systems, with relatively large core size and diluted corona, the scattering intensity in the q-range 0.03-3 $nm^{-1}$ is largely dominated by the core scattering. This observation has already been validated by others in the literature [21] with similar contrast conditions. At high pH, the data could also be fitted by a form factor of polydisperse spheres. Both treatments, at low and high pH, assume a Gaussian distribution of core sizes. From the distributions used in the adjustments we get the average core radius $R_c$ and the standard distribution $\sigma$ of the Gaussian distribution, which used in Eq. [5] to compute the mass-average weight of the micellar cores:

$$Mcore_W = d\left(\frac{4}{3}\pi R_c^3\right)\frac{1+15\sigma^{*2}+45\sigma^{*4}+15\sigma^{*6}}{1+3\sigma^{*2}} \quad \textbf{Eq. [5]}$$

where d is the density of PBA and $\sigma^*$ is equal to $\sigma/R_c$. Considering that the cores contain all the PBA from diblocks PBA-b-PAA and from homopolymer h-PBA, we computed the mass-average aggregation number $Nagg_w^{SANS}$ of diblock chains per micelles following Eq. [6]:

$$Nagg_w = \frac{Mcore_w\left(1 - X_{h-PBA}^{LC-PEAT}\left(1+X_{BA/AA}^{NMR}\right)\right)}{M_{PBAn}^{GPC}} \quad \text{Eq. [6]}$$

**Time evolution of SANS.** We used SANS to study the exchange kinetics of unimers between the micelles. For this purpose, two polymers PBA-b-PAA were synthesized with a similar targeted size of 3k-12k. One is fully hydrogenated, the other has a deuterated PBA block ($d_9$PBA-b-PAA). We checked by SANS in $D_2O$ that these two polymers form micelles of comparable size in aqueous solution. The principle of the time evolution experiment is to measure at different ages the scattering spectrum of a mixture of hydrogenated and deuterated micelles solutions. The fresh mixture is called “solution t=0”. A benchmark solution was also prepared to simulate the scattering of a solution with complete exchange between the two types of micelles. For this, we first made a cast film of the mixed polymers from THF. We had formation of spheres in the melt with cores of mixed hydrogenated and deuterated PBA. These melts were then dispersed in water to form “solution t=∞”, which corresponds to the hypothetical state of “solution t=0” after a complete exchange between the micelles. The time evolution experiment was performed in a solvent mixture $H_2O/D_2O$ of (50.5/49.5) (v/v). This mixture was chosen as a compromise. $H_2O$ amount was large enough to approach minimum scattering of micelles with mixed cores, but not low enough to limit incoherent scattering. In these conditions, the scattering of “solution t=0” was five times larger than the scattering of “solution t=∞”, which allows a good sensitivity span for detection of scattering evolution vs time.

**Cryo-TEM.** A Jeol 1200EX-120kV instrument was used at CRA, Rhodia, France. Solutions have been deposited on perforated carbon film grids. Excess of solution was blotted off in order to form a 100 nm film in the holes of the carbon film. This preparation is immediately frozen by dipping first in liquid ethane and after in liquid nitrogen. Ultra fast cooling is necessary to ensure vitrification of the solution and avoid artifact due to crystallization of the solvent or reorganization of the assemblies in solution. The frozen meniscuses in the holes are then observed by TEM.

**AFM.** We used a Nanoscope III equipped with a liquid cell. Images in water were made in tapping mode with Mikromasch tips (NSC36/AIBS, Aluminum coated Back Side). The surfaces were silicon wafers (Siltronix) cleaned by UV-$O_3$ treatment and coated with a monolayer of cationic polyelectrolyte poly(propyl methacrylate) trimethyl ammonium chloride (MAPTAC) by dipping the samples in a $10^{-2}$ wt% solution at pH 5.5. After rinsing with pure water, the samples were dipped in aqueous micellar solutions of PBA-b-PAA at pH 7 and rinsed again with pure water. Micelles are strongly adsorbed on oppositely charged surface and do not desorb upon rinsing. In order to prevent any damage of the coatings, a particular attention was paid to keep surfaces wet at all times.

### 2.4. Sample preparation.

**Cast films**: Melt polymer samples were prepared by casting films. 20 wt% solutions of diblock in THF (good solvent of the two blocks) have been deposited in Teflon molds, allowed to evaporate slowly over 7 days, and finally dried further under vacuum for 3 days.

**Aqueous solutions**: Preparation of block copolymers solutions in a selective solvent is a critical point. As previously shown in the literature, structures formed are often “out of equilibrium” [22] and may depend on the way the solutions were prepared. We used here two controlled routes to prepare aqueous solution. The first route consisted in preparing a polymer solution in a non selective solvent and exchanging the solvent via dialysis with a selective solvent (dialysis route). The solutions were then freeze-dried. The dry powder was then used to prepare the aqueous solutions at the desired concentrations. The second route consisted in preparing a cast film with organized microstructures then dissolving it in a selective solvent (cast film route). Another important parameter to adjust was the ionization $\alpha$ of the acrylic acid group. Ionization is defined as the ratio acrylic acid group that are in the salt form in the dry state. The case $\alpha = 0$ corresponds to a pure acid, whereas the case $\alpha =1$ corresponds to a pure salt (or base). The ionization was adjusted by addition of NaOH to pure acid solutions. The exact correspondence between ionization and amount of NaOH to be added was obtained experimentally by determination of the equivalence point in titration experiments.

## 3. Results and Discussion

### 3.1. Chemical analysis of components.

We present here the measurements of the main analytical characteristics of our PBA-b-PAA systems (cf. Table 1). These data will later serve for the interpretation of bulk and aqueous solution properties.

**Ethylation.** $^{13}$C NMR allows determining the fraction of acrylic acid and butyl acrylate groups that have reacted with the solvent ethanol. As shown Table 1, we can control the fraction below 5 %. This amount is larger if the reaction is run at high temperature for a longer time or if the polymers are stored in ethanol (even in a fridge at 6 °C) for a few months. It is also wise to start the reaction by the BA block because AA groups are more sensitive to esterification. Starting by the PBA block decreases the time during which AA is heated in the solvent. It has been shown that the presence of ethyl acrylate hydrophobic moieties in the PAA block can act as stickers between PAA chains in aqueous solutions [23]. Indeed, we have observed by AFM and light scattering the formation of aggregates of micelles in aqueous solution and a general decrease of solubility of the samples with higher ethylation degrees. This problem can in principle be totally avoided by the use of another solvent, that has no alcohol function [24]. However, all other solvents have higher risks associated with their use at industrial scale and we show here that ethylation can be limited to a level low enough to avoid changes of general properties.

**$M_w$ and polydispersity.** We present on Figure 2 the GPC data of a PBA-b-PAA 4k-20k for aliquots of the first block and for the whole methylated diblock. The Mw of the first block and the total chain are respectively measured at 4.5k and 30k. These values are very close to the targeted molecular weights of 4k and 24k targeted upon synthesis. As an alternative way to determine the molecular weight of the second block, we used the mass ratio of the diblock component of the mixture $X_{BA/AA}$ . The molecular weight of the second block was taken as the molecular weight of the first block (determined by GPC), multiplied by the ratio $X_{BA/AA}$. The total $M_W$ found by this method is 20k which is quite consistent with the targeted and GPC values. It has the advantage to take into account the correction due to the presence of homopolymers h-PAA and h-PBA. The values of $M_W$ of second block on **Table 1** are determined by this method. On **Table 1**, the values of $M_w$ are consistent with the targeted values $M_t$ estimated by Eq. [1], whereas $M_n$ values are always smaller. These results have been shown to arise directly from the synthesis mechanism, which is characterized by a relatively slow exchange of the controlling agents ($C_{Xa} \approx 2.7$) and a high transfer to solvent constant ($C_s \approx 4.7\ 10^{-3}$) [13]. On one hand, $I_p$ values of 1.4-2 are a significative improvement as compared to values obtained with uncontrolled radical polymerization (typically 3 or more). However, GPC spectrum illustrates clearly that $I_p$ values of 1.4-2 correspond to chain length distributions spreading over 2 decades. As shown in this paper, this characteristic has consequences on the sample properties in solution.

**Homopolymer h-PBA.** The mass fraction $X_{PBA/tot}^{LC-PEAT}$ of homopolymer h-PBA in the samples have been determined by LC-PEAT (Table 1). They are always around 5-10 wt% of the total polymer mass. This relatively large amount was attributed to transfer to solvent events [13]. In the end, the presence of dead chains of h-PBA hardly affects the properties of the samples. In the melt state or in aqueous solution, the h-PBA is trapped in the cores of PBA-b-PAA micro-phase separated structures. The main effect is a core swelling. Measured value of h-PBA content will be taken into account to interpret the structural characteristics of objects.

**Homopolymer h-PAA**: The mass fraction $X_{h-PAA}^{CE}$ of homopolymer h-PAA in the samples have been determined by Capillary Electrophoresis (Table 1). We have found typical values of 30 wt% versus the total polymer mass. This large amount was attributed to the high polymerization rate of PAA as compared to the exchange constant of controlling exchange xanthate $C_{Xa}$ [13].

### 3.2. Physical analysis of components.

We present here the measurements of the main physical characteristics of our PBA-b-PAA systems. These data will later serve for the interpretation of self assembling properties in solution.

**Interfacial tension h-PBA/Water**. Interfacial tension h-PBA/Water was measured by drop shape analysis of an h-PBA sessile drop immersed in pure water. h-PBA density is close to water density, which makes the drop shape analysis method sensible to a small variations of h-PBA densities. The exact value of h-PBA density depends on the end group type and on the molar mass of the chains (by a dilution effect of the end groups). h-PBA densities were therefore measured for each sample used in interfacial tension experiments. Most of interfacial tension experiments were performed with a high molar mass sample ($M_w$ = 60k, $I_p$=3, secondary standard, Aldrich 181412) that we precipitated from toluene in methanol in order to eliminate the smallest chains. Indeed, we observed that the presence of small chains induced a slow decrease of the interfacial tension over several hours. Finally, for the purified sample, we measured a density $d_{PBA} = 1.02 \pm 0.01$ g.cm$^{-3}$ and we found a stable interfacial tension value value $\gamma_{PBA/H2O} = 20 \pm 2$ mN/m. This value is consistent with indirect experimental determination found in the literature [25].

**Flory parameter**. **Micro-phase separation**. PBA is a liquid at room temperature and PAA is a solid. The melt films are all solid with a more rubbery aspect when the ratio $X_{BA/AA}^{diblock}$ is larger. In any case, the structures in the melt state are physically frozen. Therefore, melt structures are not strictly speaking equilibrium structure of pure polymer, but rather frozen structures remnant of a micro-phase separation that occurred sometime during the drying process, when enough solvent was still present to allow polymer chain movement. Note that there is no better way to approach thermodynamically stable state for melts of PBA-b-PAA diblock because a temperature annealing above $T_g$ of PAA burns the samples. This is why the approach of solvent evaporation is used in literature [26,27]. The topologies of the structures are determined by SAXS.

Figure 5 presents examples of SAXS spectra for PBA-b-PAA 6k-24k and PBA-b-PAA 3k-4k. For PBA-b-PAA 6k-24k, a clear $q^{-4}$ dependence appears at high q, which is a typical Porod regime of scattering objects with sharp interfaces. The overall spectrum can be adjusted by a model of polydisperse spheres with an average core radius $R_c$ of 114 Å and a normalized standard deviation $\sigma$ of 0.18. Other adjustments parameters for other melt samples with spherical topologies are reported on Table 2. The table reports also the aggregation numbers of chains PBA-b-PAA per sphere calculated by two methods: $N_{BA}$ is determined by simply dividing the volume of a sphere of radius $R_c$ by the volume of a PBA block, whereas $Nagg_w$ is determined using Eq. [6] and represents the weight average aggregation number of micelles as determined from the SAXS. The two calculations lead to slightly different values, which illustrates that aggregation numbers have to be defined cautiously for polydisperse systems. For PBA-b-PAA 3k-4k, the SAXS spectrum of

Figure 5 is typical of a lamellar phase. The two peaks correspond to order 1 and 3. The even peaks are absent, which means that the PBA and PAA layers have a similar thickness. The repeating distance for PBA-b-PAA 3k-4k is found at 210 Å. The overall phase diagram for the melt state versus the volume fraction of BA $f_{BA}$ and the diblock length in number of monomers N is reported on Figure 6. This diagram is consistent with self consistent field theory (SCFT) predictions [28] and experimental findings on similar systems. On the basis of the SCFT, and using the order-disorder boundary, one can extract an estimation of the Flory parameter $\chi_{PBA/PAA}$ between PBA and PAA of 0.25. This system is therefore a strongly segregative system. As already reported in the literature [26], we also confirm here that the high polydispersity of CRP diblocks copolymers are not prohibitive for the formation of long range ordered systems.

### 3.3. Dispersions in water

**Preparation route dependency**. By the cast film route, samples dispersed in pure water formed solutions with a natural pH around 3. The initial topologies present were modified in some cases upon dispersion in water, as a result of the swelling of the PAA corona. For instance, PBA-b-PAA 3k-4k which forms lamellar structure in the melt state and transforms into cylinders in aqueous solution at pH 3 (Figure 7). By the dialysis route, all samples studied formed spherical micelles. It is interesting to remark that PBA-b-PAA 3K-4k solutions present very different states vs the way of preparation, spheres via dialysis route and cylinders via cast film route. This preparation route dependency is a first indication that self assembling of PBA-b-PAA in aqueous solution is not in a thermodynamically stable state. In what follows, we investigate further these self assembling characteristics by focusing on the case of spherical micelles in aqueous solution.

#### 3.3.1 Structural characterization of aggregates.

**Core characterization.** The cryo-TEM picture of PBA-b-PAA 3k-4k solution at ionization $\alpha$=0 Figure 7 shows a population of spheres of mean radius 18 nm that corresponds to PBA cores. Figure 8 shows the SANS and SAXS scattering of the same PBA-b-PAA 3k-4k solution at ionization $\alpha$=0 in $D_2O$. The first minimum of the form factor corresponds to a radius $R = 4.52/q_{min} = 18 \pm 1$nm, which is consistent with the core size observed by cryo-TEM. The whole spectra could be adjusted by a Pedersen model for both SANS and SAXS data (**Figure 8**). The same core-corona characteristics ($R_c$ = 16.6 nm, $N_{agg}$ = 2600 , $l_K$ = 0.4 nm) were used for both Pedersen fits. A contrast ratio between the core and the corona was taken at 5 for X-ray and 2 for neutrons, which corresponds to theoretical contrast values. The fact that the same structural model fits scattering data in different contrast conditions is a proof of accuracy of the model description. This set of data on PBA-b-PAA 3k-4k shows that core size values can be determined precisely. We have checked on numerous samples that the core scattering is dominant over the corona scattering in the q range 0.003-0.3 $Å^{-1}$ for SANS data in $D_2O$. Significant contribution of the corona scattering is only detected for samples with a deuterated core (the contrast between the core and the solvent is small) or for crew-cut micelles with non ionized corona (the corona is dense enough to contribute to the scattering). Hence, in most cases, precise information about the core size was easily accessed from SANS data and polydisperse spheres model adjustments. This is especially valid in the regime of charged and extended corona. The normalized standard deviation of the core size distributions were found around 0.13, which is a typical value for micelles of diblock synthesized by either ionic [29] or controlled radical polymerization [19,22,23,26]. The dependence of the core size versus the molecular weight of the PBA block is reported on insert of Figure 8 for a series of diblocks with a constant PBA/PAA ratio of 0.25. The solid lines correspond to theoretical predictions [30,31] made for star like micelles with a corona in good solvent: the core size $R_c$ is expected to vary with a power law of 2/3 versus the degree of polymerization of the first block. Data and theory show a relative consistency, but this may just be a coincidence since the range of molecular weights investigated is narrow and micelles are not in a thermodynamic stable state. More interesting, the absolute core sizes extracted from SANS measurements are quite large. They are almost equal to the maximum size authorized by the stretched length of the inner PBA block $SL_{PBA}$ (cf. Table 2). Such puzzling results are usual in the literature with samples made by CRP [26] without a clear explanation. Several parameters can favor the formation of large cores. First, the PBA blocks are quite polydisperse, as pointed in the analysis part. The maximum core size is not limited by the average size of the PBA blocks because larger blocks are present in solution. Also, there is homopolymer h-PBA in the samples. This h-PBA, being very hydrophobic, is necessarily trapped in the micelles cores. This effect tends obviously to swell the cores. Let us assume that the fraction of h-PBA in each micelle core is equal to the average fraction of h-PBA in the sample, and that all h-PBA is located in the center of the micelles (like an emulsion). We can then use the h-PBA content determined in the analysis part and calculate the thickness R* of the PBA corresponding to PBA-b-PAA chains that surrounds h-PBA spheres. Calculated values R* are reported on Table 2. R* values are significantly smaller than the experimental value $R_c$. h-PBA content plays therefore a major role in the swelling of micelles cores, which explains why controlled radical samples always lead to abnormal large cores. Another argument given in the literature is that the core may contain a mixture of the two blocks. This argument does not stand for PBA-b-PAA, due to the high Flory parameter $\chi_{PBA/PAA} \approx 0.25$ and the high hydrophilicity of ionized PAA.

**Corona characterization.** DLS and CONTIN analysis permit a direct determination of hydrodynamic radius $R_h$ of micelles. In a few cases, the DLS data could be compared to direct AFM images (for the largest micelles), and to SANS data (for micelles with deuterated cores). With AFM, the diameter of 120 nm of PBA-b-PAA 6k-24k micelles observed in **Figure 9** is comparable to the diameter measured by light scattering, i.e. 140 nm. With SANS, corona information could be extracted on a deuterated sample $d_9$PBA-b-PAA 3k-12k. With this deuterated sample, the low contrast between the core $d_9$PBA and $D_2O$ permits to enhance the scattering of the corona as compared to the scattering of the core, which is largely dominating in fully hydrogenated samples. **Figure 10** presents the scattering of a $d_9$PBA-b-PAA 3k-12k 1.5 wt% at $\alpha$=0 and $\alpha$=1 in $D_2O$. The data at $\alpha$=0 have been adjusted with a Daoud Cotton model ($R_c$=7.7nm, $R_h$=20nm, n=–4/3, and $\rho_c$/ $\rho_1$=3.3). The density profile used for the fit is given in the insert of **Figure 10**. The data at $\alpha$=1 have been adjusted with an Urchin model ($R_c$=7.7 nm, $L_{rod}$=50nm, $N_{agg}$=300). Comparatively to the scattering of a fully hydrogenated PBA-b-PAA 3k-b-12k sample, we observed that the absolute intensity at low q is divided by 10, whereas the core size remains similar for the two samples. For a fully hydrogenated sample, the low q scattering is completely dominated by the core scattering, which explains it is hardly affected by ionization (the cores are hardly affected in most cases) and why it decreases sharply when the contrast of the core is decreased. For sample with deuterated cores, the low q scattering is a combination of core and corona scattering contributions. Although the core is hardly affected by ionization, the corona stretching is obviously affected by ionization which is detected in the low q scattering. The q-range investigated is only large enough to grasp the whole scattering at zero angle for the case of low ionization (small collapsed micelles). For this case, **Figure 10** presents the scattering data in two contrast conditions, in pure $D_2O$ and in a mixture $D_2O/H_2O$ 82/18 (v/v). The scattering spectra could be adjusted by a Guinier model with a $R_g$ of 16 nm, or an effective radius of 20 nm. More precisely, a core-corona model with a core of 7.7 nm and a total micelle radius of around 20 nm worked for both contrast conditions. These values are in good agreement with the hydrodynamic radius $R_h$ = 22 nm measured by DLS. In the end, DLS, AFM and SANS give consistent information on the micelles corona extensions. SANS is not adapted to the case of very large micelles in the ionized state, and AFM is only adapted to the largest micelles made of large molar mass diblocks. In the following, all corona information are inferred from DLS measurements.

**Influence of concentration, addition of salt and ionization.** The methods exposed for a precise determination of core sizes and corona sizes are now used to investigate the influence of concentration, addition of salt and ionization. On, the corona of PBA-b-PAA 3k-12k micelles undergo a stretching by a factor 3 between the neutral state (pH=2) and the fully charged state (pH=8). In the charged state, with a core of $R_c$=7 nm and a hydrodynamic of $R_h$ = 51 nm, the micelles look like stars with highly stretched arms. Note that the extension of arms of 44 nm is close to the maximum extension of the average stretched chain of PAA 12k. Indeed, a PAA 12k chain of 166 monomers has a linear length of 42 nm. Obviously, the chains in the corona are not completely linear. We have seen that the polydispersity of the blocks is at best of 1.5. The corona size as deduced from the hydrodynamic radius is therefore sensitive to the extension of chains that are longer than 12k. In Figure 11, the diameter of the corona shows a pronounced decrease from 48 nm to 33 nm upon addition of salt. This collapse occurs over several decades of salt concentration. For a planar osmotic brush, it is well established that the extension of the brush [32] decreases sharply with salt concentration above a threshold value that corresponds to the internal salt concentration of the brush. Here, the geometry is spherical so that the corona concentration decreases strongly with the radial distance. The threshold value is therefore spread on a large range of concentrations. Also, the polydispersity of corona chains is very high and the small portion of longest PAA chains protruding far away from the micelles center have practically no neighbor. These portions are extremely sensitive to very small addition of salt, which explains why the corona collapse starts at salt concentrations as low as $10^{-4}$ M. We then investigated the dependence of the core size vs ionization and salt concentration by SANS. We finally found that the core sizes are hardly affected by polymer concentration and salt addition, which is in complete disagreement with the behavior of a system at thermodynamic equilibrium. Conversely, upon increasing of pH, i.e. upon charging of the corona, a decrease of the core size, although limited, was detected by scattering experiments and cryo-TEM pictures. This decrease is important and therefore easier to detect with symmetric diblocks, whereas it is much weaker with asymmetric diblocks. The cryo-TEM pictures (Figure 13) show a transition from large and relatively monodisperse spheres to smaller and more polydisperse spheres. The scattering experiments (Figure 12) present a decrease by a factor 10 of the intensity at low q upon ionization of the micelles. As the total concentration is constant and as the contrast is only marginally modified by the pH change, the scattering decrease is necessarily due to decrease in the size of the scattering objects. More precisely, the adjustments of a series of scattering data at different ionization ratios has permitted to follow the decrease of the average core size versus the ionization (insert of Figure 12). We found a gradual decrease of core sizes from 18 to 13 nm between ionization $\alpha$=0 and $\alpha$=1. At first, this result seems in qualitative agreement with the behavior expected with a system at equilibrium. Indeed, charged diblocks have a bulkier hydrophilic block. They are then expected to form objects of smaller aggregation number and higher radius of curvature. However, the structural changes of the cores observed with pH are not reversible. The object observed at $\alpha$=1 kept the same aggregation number when the pH was lowered again. The behavior with pH is then strongly hysteretic which is another a sign of non equilibrium.

### 3.3.2 Critical micelle concentration

We challenge now further the question of reversibility of PBA-b-PAA micellization in water. From a thermodynamic point of view, the formation of micelles appears above a minimum concentration of unimers, known as the CMC. For a system at thermodynamic equilibrium, there is only unimers in solution below the CMC and there are both micelles and unimers above the CMC. Several techniques permit to determine CMC values. The surface tension of a solution vs concentration decreases below the CMC (following Gibbs equation) until a break point that corresponds to the CMC. The surface tension is then stable vs concentration above the CMC. Figure 14 shows the surface tension measurements of PBA-b-PAA 1k-4k solutions versus concentration. The solutions are prepared by dilutions of the highest polymer concentration solution. of Hollow points correspond to measurements taken one hour after formation of a clean interface. This plot shows an apparent CMC around 0.03 wt%, which is consistent with data found in the literature [33,34,35] obtained with an automatic device that measures surface tensions one hour after each increment of concentrations. However, we found that adsorption kinetics could take much longer than one hour, especially at low concentrations. On Figure 14 the dark data points correspond to equilibrated surface tension values. In this plot, no CMC is detectable down to a concentration of 0.003 wt%. Investigations at lower concentrations became prohibitively time consuming by this technique (equilibration times over 24 hours) and this concentration limit is even higher with diblocks of longer PBA blocks, whose adsorption kinetics is even slower. We also used static light scattering measurements to probe the presence of micelles in solution at different concentrations. For all concentrations, the scattering was always resulting micelles and the total scattering was strictly linear with concentration above the concentration for which the signal becomes so low that it is lost in the background scattering of water. This means that no CMC can be detected by SLS in the range of concentration limited by water background scattering (down to $10^{-4}$ w%). We finally tried to use the technique of pyrene fluorescence shift to detect the presence of micelles. We found a transition between pyrene in water and pyrene in PBA cores that happened at the same mass concentration for all PBA-b-PAA samples. However, we determined that this transition did not correspond to a CMC in solution. Indeed, we proved indeed that this apparent transition was rather resulting from a partition effect of pyrene between water and PBA. Although pyrene is much more soluble in PBA than in water, at very high dilutions, the partition of pyrene between a tiny amount of PBA and a huge amount of water turns to the advantage of water. This assessment was quantitatively proven by the experimental determination of the partition constant of pyrene between pure water and pure h-PBA ($K_{py} \approx 7000$). Finally, from all these measurements, it has not been possible to determine the values of the CMC of PBA-b-PAA systems because they are extremely low. The question is then: is the self assembly of PBA-b-PAA reversible with extremely low CMC's, or is this self assembly irreversible (or out of equilibrium)and the concept of CMC (which is a thermodynamic concept) is irrelevant. In the reversible case, there is a transition vs concentration between a regime with and without micelles, whereas in the irreversible case, there are micelles at any concentration in solution. Previous observations of non equilibrium behavior for PBA-b-PAA systems tend to validate the later conclusion.

#### 3.3.3 A double population of diblocks in solution

A direct consequence of extremely low CMC is that the concentration of free unimers in solution is supposed to be approximately equal to the CMC, i.e. extremely low. This has been checked by Capillary Electrophoresis experiments (Figure 4). Peak (1) at –33 $10^{-5}$ $cm^2.V^{-1}.s^{-1}$ has been attributed to h-PAA by addition of h-PAA to the samples, which increases the intensity of peak (1). The double peak at –28 $10^{-5}$ $cm^2.V^{-1}.s^{-1}$ has been attributed to the diblock PBA-b-PAA by addition of surfactant Brij 35, which affects the position and shape of this doublet (Figure 4). It is indeed expected for diblocks to interact with surfactant via their hydrophobic PBA moiety. Networks of dextran have then been used to separate the species vs their size. This has permitted to determine that the intense peak (3) at $\mu_{ep}$ -27 $10^{-5}$ $cm^2V^{-1}s^{-1}$ corresponds to large objects, i.e. micelles of PBA-b-PAA, whereas the smaller peak (2) at $\mu_{ep}$ -28.7 $10^{-5}$ $cm^2V^{-1}s^{-1}$ corresponds to small objects, i.e. free unimers of PBA-b-PAA [38,39]. It is clear that peak (2) (unimers) is far from negligible as compared to peak (3) (micelles). This sounds contradictory with the result of extremely low CMC value. We then measured the dependence of the intensities of peak (2) and (3) versus sample concentration (Figure 4). Both variations are linear. In the scope of CMC thermodynamics, only the micelle population grows with concentration above CMC and unimer content remains constant around the CMC value. The large unimer population does NOT fit with a simple CMC model. Our explanation is that the large polydispersities of PBA blocks have to be taken into account. The idea is that the population of free unimers observed correspond to diblock chains with very short PBA blocks. These chains, corresponding to peak (2) in CE, have a very high CMC, if any. They do not associate and their quantity varies linearly with concentration. The population of associated diblock chains corresponds to diblocks with long PBA blocks, which have extremely low CMC and form micelles at all concentrations. This micelles amount varies also linearly with concentration. Whereas in a system at equilibrium, micelles and unimers populations are made of non distinguishable chains that exchange between the two populations, in our case, the two populations are made of physically different chains that behave differently in solution. This latter point has been checked by separating physically the two populations via *preparative* aqueous GPC. After separation, the micelle population was re-injected in Capillary Electrophoresis and a single peak was observed. To conclude this CMC discussion, one should note that the presence of free unimers in solution of amphiphilic diblocks with low CMC is not specific to radical polymerization samples. Free unimers have also been observed in solution of Poly(tertbutyl Styrene)–Poly(Styrene Sulfonate) [6] and Poly(Styrene)-Poly(Acrylic Acid) [34] diblocks in water, although these systems had been synthesized by anionic polymerization and form frozen core objects.

### 3.3.3 Reversibility and dynamics of self assembly.

Several arguments have shown that PBA-b-PAA samples in aqueous solution are not at thermodynamic equilibrium, (i) the topology of the aggregates depends on the preparation route, (ii) the micelles aggregation number does not depend on concentration and salt addition but present an hysteresis with ionization cycles, and (iii) the CMC are not detectable. This latter characteristic has been observed in numerous systems of diblocks in aqueous solutions like PS-b-PAA or PtBSt-PSS [19,20,22,27,37,38]. In these examples, the core was made of a polymer with a high glass transition. The irreversibility could then be attributed to the glassiness of the cores. However, with PBA cores ($T_g$ = -54 °C), the micelles cores are liquid at room temperature. The cores are not physically frozen and the exchange of unimers between micelles is in principle possible. Indeed, some reorganizations of micelles upon increasing the ionization of the corona have been observed, which is impossible with glassy cores. The equilibrium state of PBA-b-PAA solution may just be a question of time. Therefore, we have attempted a measurement of the timescale needed in PBA-b-PAA systems for exchanges to take place.

**Equilibrium time scale determination [41].** We have used two diblock samples synthesized with similar block sizes and different deuteration degrees, $d_9$PBA-b-PAA 3k-12k and PBA-b-PAA 3k-12k (cf. part 2.3). The scattering of mixtures of the two types of micelles versus time as well as the benchmark "solution t=∞" are reported on Figure 15. At time t=0 the scattering at low q is 5 times larger than the benchmark curve. We have measured the evolution after a maximum of 6 months and no detectable change was observed. The kinetics of exchange of unimers of PBA-b-PAA 3k-12k is therefore negligible at the time scale of several months. The system is kinetically frozen, i.e. there is a very high energy barrier preventing the exchange of unimers.

**Fractionation.** Based on this result, the reorganization of the structure of micelles cores observed upon corona charging can not occur via exchange of unimers. It must occur by a mechanism internal to each micelle. In other words, the micelles fractionate under the increasing pressure of the charged corona. Upon charging, the corona swells and forces the micelles to fractionate. Obviously, this mechanism does not lead to a thermodynamically stable state, which explains that the polydispersity of micelles size increases (Figure 13). Note also that upon neutralization of PAA, the constraint in the corona decreases, which, at thermal equilibrium, would give the opportunity to the micelles to grow to larger sizes. However, any structural change can only happen within a single micelle and a micelle growth implies the arrival of new material from other micelles. This incompatibility explains why the cores size remains the same upon neutralization of the PAA corona or addition of salt; i.e. fractionation explains the hysteresis of size observed versus an ionisation cycle. These explanations can now be compared to other systems such as poly(styrene)-b-poly(acrylic acid) diblocks (PS-b-PAA) [42].

The structures formed in the melt (spheres, cylinder and lamellae) for different PS-b-PAA samples with ratio PS/PAA between 0.25 and 0.5 kept their topological structure at room temperature when dispersed in water even after ionizing the corona. The high glass transition temperature of polystyrene ($T_g$ around 100°C) imposed the cores of polystyrene to be frozen at room temperature. After several hours at 90°C, structural evolutions were observed from lamellae and cylinders to spheres. A fractionation of cylinders and spheres under ionization of the corona was also observed. The behavior of PS-b-PAA close to the Tg of PS is therefore very similar to the behavior of PBA-b-PAA at room temperature, i.e. above the $T_g$ of PBA (-54°C).

**3.3.4. Interfacial tension as a control parameter for reversibility.**

We will explain here why PBA-b-PAA is a kinetically frozen system in aqueous solution by a simple calculation of the CMC of the systems [43]. The interfacial contribution of the free energy of PBA-b-PAA chains in unimer form and in aggregates of N chains can then be written as Eq. [7]:

$$\mu_{unimer}^{0}=4\pi R_{uni}^{2}\gamma_{PBA/H2O} \quad \text{and} \quad \mu_{N}^{0}=\frac{4\pi R_{agg}^{2}\gamma_{PBA/H2O}}{N_{agg}} \qquad \textbf{Eq. [7]}$$

where Ragg is the core radius of the aggregate (accessible experimentally by SANS) and $R_{uni}$ is the radius of the PBA block. $R_{uni}$ is calculated based on the mass of the PBA block, the density of PBA $d_{PBA}$ measured at 1.02 ± 0.01 g.cm$^{-3}$ and he assumption that the block forms a spherical drop. The contribution of PAA to the chemical potential is neglected here. It is indeed easy to show that this contribution is negligible when the interfacial tension between the core of the micelles and the solvent is high enough [44]. The CMC can then be written as:

$$CMC=\exp\left(-(\mu_{unimer}^{0}-\mu_{N}^{0})/kT\right) \qquad \text{Eq. [8]}$$

The aggregation number being very large in our systems, the free energy of chain in the aggregates can be neglected. This CMC calculation correspond therefore to the calculation of a solubility limit of PBA block in water. Eq. [8] can be simplified into:

$$CMC\approx\frac{100M_{BA}(X_{BA/AA})}{vN_a(1+X_{BA/AA})}e^{-\pi^{1/3}(3/4nv)^{2/3}\gamma/kT} \qquad \text{Eq. [9]}$$

where the CMC is in wt% of sample PBA-b-PAA, $M_{BA}$ is the molar mass of a monomer BA, $X_{BA/AA}$ is the mass ratio of BA over AA in a hypothetic monodisperse diblock sample considered in this calculation, v is the molecular volume of a BA monomer, $N_a$ is the Avogadro number, n is the polymerization number of the PBA block There is no adjustable parameter since interfacial tension $\gamma_{PBA/H2O}$ has been measured at 20 mN/m by sessile drop shape analysis. Figure 16 presents the calculated CMC values vs the molar mass of the PBA block for a series of PBA-b-PAA samples with $X_{BA/AA}$ of 0.25. In the range of PBA molar masses investigated, between 2k and 6k (cf. Table 1), the calculated CMC values are very low, between $10^{-3}$ wt% and $10^{-7}$ wt%. This shows that the value of $\gamma_{PBA/H2O}$ is high enough to impose extremely low CMC values, a high energy cost for extraction of a unimer from a micelle, and the slow kinetics of exchange of unimers between micelles. Figure 16 shows also that the CMC varies very fast with the molecular weight. This characteristic is very important to explain further the presence of the two distinct populations in solution, the one of free unimers and the one of aggregated unimers. One point is a puzzling at first. These two populations differ only by their PBA block length which confers to them very different and independent behaviors. At the same time, the distribution of PBA blocks (cf. **Figure 2**) is clearly not bi-disperse. Our interpretation is that the sharp dependence of the CMC vs the PBA length implies a sharp transition of aggregation behavior. Two apparent populations appear even with diblocks that have a continuous distribution of PBA block lengths.

All our interpretations can be extended to other systems like PS-b-PAA [26], for which the irreversibility of the aggregation is sometimes attributed to the glassiness of the cores that makes the micelles physically frozen. The interfacial energy PS/$H_2O$ has been estimated around 35 mN/m which implies that PS-b-PAA micelles in water are also kinetically frozen, even more than PBA-b-PAA. For PS-PAA diblocks with small PS blocks and forming micelles with small cores, the glassiness in confined environment is questionable. The PS-b-PAA samples always form irreversible aggregates in water and the interfacial tension argument is able to explain it alone.

**4. Summary.**

We have presented a link between an in depth chemical and physical characterization of amphiphilic diblock copolymers samples PBA-b-PAA and their self-assembling properties in aqueous solutions. The samples were synthesized by controlled radical polymerisation MADIX®. Despite imperfections of the products, these PBA-b-PAA samples present the general characteristics expected for ideal diblock systems: they self-assemble in the melt state and in aqueous solution with different topologies depending on the monomer ratio BA/AA. Divergences from ideal behaviour, which are often overlooked in the literature, could here be explained with the characterization data. The sizes of the cores were found generally large as compared to the contour length of the PBA block. The high Flory parameter $\chi_{PBA/PAA}$ value, which we determined at 0.25, excludes the possibility that PAA swells the cores. The homopolymer h-PBA content, which was determined by LC-PEAT, was found large enough to explain quantitatively the apparent swelling of the cores. The polydispersity of the blocks, determined between 1.5 and 2.5 have been shown to be responsible for the large values of hydrodynamic radii, as compared to the average size of the PAA blocks. An investigation by capillary electrophoresis has shown the presence of a large population of free unimers in solution. This opened the question of the equilibrium between chains "free in solution" and chains "aggregated in micelles". The CMC of the systems have been found extremely small, which is in accordance with the high value of interfacial tension between PBA and water that we determined at 20 mN/m, but not with the presence of a large population of free unimers. In fact, preparative GPC analysis and capillary electrophoresis CE experiments have shown that the population of chains "free in solution" corresponds to diblock chains with very small PBA blocks. These chains never aggregate and are not in equilibrium with the micelles. Conversely, the micelles have then been shown to be kinetically frozen as a direct consequence of the high interfacial tension value $\gamma_{PBA/H2O}$ . Hence, the structural state of the solution is strongly preparation route dependent, and also structural changes of aggregates vs concentration, ionic strength and ionisation are limited or null. In this context, the only structural reorganizations observed have been attributed to a fractionation mechanism of micelles. In the end, this paper has proposed an in-depth link between chemical analysis and self assembly properties of diblock copolymers PBA-b-PAA. It has also elucidated the peculiar behavior of diblock copolymer solutions at the edge of self-assembly reversibility.

For future work, this family of MADIX® PBA-b-PAA sample, which has the main characteristics of a pure diblock system, is valuable for both industrial and academic purposes. The next step will be to investigate their adsorption properties. PBA-b-PAA samples have already proven great efficiencies for emulsion and reverse emulsion stabilization. Also, it appears interesting to modify the block PBA in order to reduce its interfacial tension with water and change the aggregation properties in water in order to come closer to reversibility.

Thanks to : M. Destarac and G. Lizzaraga for advises on the synthesis part, C. Bauer, H. Mauermann and D. Radtke for active participation in the analysis measurements and preparative GPC experiments, L. Porcar from NIST and F. Cousin from LLB for help on the scattering runs, A. Checco and P. Guenoun for technical help with AFM experiments, and also M. Airiau and A. Vacher for help with cryo-TEM experiments.

## LIST OF TABLES

Table 1: Chemical analysis results. $M_n$, $M_w$, are the number average and mass average molar masses, $I_p$ is the index of polydispersity, $X_{BA/AA}^{NMR}$ is the mass ratio of BA versus AA in a sample, $r_{EA}^{NMR}$ is the number fraction of ethyl acrylate groups in a sample, $X_{h-PBA}^{LC-PEAT}$ is the mass fraction of homopolymer h-PBA in a sample, $X_{h-PAA}^{CE}$ is the mass fraction of homopolymer h-PAA in a samples, and $X_{BA/AA}$ is the mass ratio between BA and AA of the diblock component of a sample.

| | **GPC** | | | **NMR** | | | **LC-PEAT** | **CE** | |
|---|---|---|---|---|---|---|---|---|---|
| **Sample name (with $M_t$ values for each block)** | **$M_{nBA}$** | **$M_{wBA}$** | **$I_{p\ BA}$** | **$M_{w\ AA}$** | $X_{BA/AA}^{NMR}$ | $r_{EA}^{NMR}$ | $X_{h-PBA}^{LC-PEAT}$ | $X_{h-PAA}^{CE}$ | $X_{BA/AA}$ |
| PBA*-b-PAA "3k-12k" | 3620 | 5430 | 1.5 | 22600 | 0.24 | 0.04 | 0.06 | 0.49 | 0.42 |
| PBA*-b-PAA "4k-20k" | 2990 | 4545 | 1.52 | 18937 | 0.24 | 0.01 | 0.04 | 0.3 | 0.30 |
| PBA*-b-PAA "6k-24k" | 2600 | 6420 | 2.47 | 29200 | 0.22 | 0.04 | 0.09 | 0.51 | 0.29 |
| $d_9$PBA*-b-PAA "3k-12k" | | - | - | - | 0.24 | 0.04 | 0.02-0.06 | 0.34 | 0.28 |
| | **$M_{nAA}$** | **$M_{wAA}$** | **$I_{p\ AA}$** | **$M_{wBA}$** | | | | | |
| PBA-b-PAA* "1k-4k" | 3670 | 8850 | 2.41 | 2300 | 0.26 | 0.09 | 0.08 | 0.35 | 0.28 |
| PBA-b-PAA* "3k-4k" | 3800 | 10430 | 2.72 | 7200 | 0.69 | 0.07 | 0.09-0.2 | 0.25 | 0.61 |

**Table 2: SAXS results on cast films; Rc is the core radius, $N_{BA}$ is the aggregation number determined by dividing the volumes of a sphere of radius $R_c$ and of a PBA block, and $Nagg_w$ is and represents the weight average aggregation number of micelles determined using Eq. [6].**

| | $R_c$ (nm) | σ | $N_{BA}$ | $Nagg_w$ | $S_{LA}$ (nm) | R*(nm) |
|---|---|---|---|---|---|---|
| 6k-24k | 11.4 | 0.14 | 580 | 930 | 11.7 | 2.1 |
| 4k-20k | 9.0 | 0.17 | 400 | 690 | 7.8 | 3.5 |
| 3k-12k (d) | 7.7 | 0.17 | 380 | 640 | 5.9 | 2.3 |
| 1k-4k | 4.1 | 0.25 | 80 | 205 | 2.0 | 1.0 |

**Table 3: dn/dc values at $\alpha$=0 and 8 for the PBA-b-PAA samples studied in aqueous solution.($\lambda$=654 nm)**

| Sample | 1k-4k | 3k-12k | 6k24k |
|---|---|---|---|
| dn/dc (pH=2) | 0.131 | 0.133 | 0.136 |
| dn/dc (pH=9) | 0.177 | 0.181 | 0.160 |

**Table 4: Elements characteristics for X-ray and neutron scattering; $\rho_e$ is the calculated electronic density, $\rho$ is the scattering length density , $\Delta b_{D20}$ is the difference of scattering length density with D2O.**

| Polymer | $\rho_e$ (ē/Å$^3$) | $\rho$ (.10$^9$ cm$^{-2}$) | $\Delta b_{D20}$ (.10$^9$ cm$^{-2}$/g) |
|---|---|---|---|
| PBA | 0,336 ± 0,002 | 6,35 | -56,0 |
| $d_9$PBA | - | 51,4 | -11,0 |
| PAA-H | 0,480 ± 0,14 | 21,0 | -28,1 |
| PAA-Na | 0,854 ± 0,52 | 42,7 | -7,46 |
| $d_3$-PAA-Na | - | 98,4 | 12,2 |
| PAA-Cs | 1,059 ± 0,72 | - | - |

**References**

[1] M.F. Cunningham, Progress in Polymer Science 27 (2002) 1039.
[2] J. Chiefari, Y.K. Chong, F. Ercole, J. Krstina, J. Jeffery, T.P.T. Le, R.T.A. Mayadunne, G.F. Meijs, C.L. Moad, G. Moad, E. Rizzardo, S.H. Thang, Macromolecules 31 (1998) 5559.
[3] J.S. Wang, K. Matyjaszewski, Macromolecules 23 (1995) 7901.
[4] V. Coessens, T. Pintauer, K. Matyjaszewski, Progress Polymer Science 26 (2001) 337.
[5] R. Xu, Y. Hu, M.A. Winnik, G. Reiss, M.D. Croucher, J. Chromatogr. 547 (1991) 434.
[6] H. Cottet, P. Gareil, P. Guenoun, M. Muller, M. Delsanti, P. Lixon, J.W. Mays, J. Yang, J. Chromatogr. A 939 (2001) 109.
[7] J. Jansson, K. Schillén, G. Olofson, R. Cardoso da Silva, W. Loh , J. Phys. Chem. B 108 (2004) 82.
[8] S.L. Nolan, R.J. Phillips, P.M. Cotts, S.R. Dungan , J. Colloid Interface Sci. 191 (1997) 291.
[9] G. Gallet, S. Carroccio, P. Rizzarelli, S. Karlsson , Polymer 43 (2001) 1081.
[10] G. Marinov, B. Michels, R. Zana, Langmuir 14 (1998) 2639.
[11] P. Linse, T.A. Hatton, Langmuir 13 (1997) 4066.
[12] MAcromolecular Design via Interchange of Xanthate, Rhodia Patent WO 9858974.
[13] M. Jacquin, P. Muller, C. Bauer, G. Lizaragua, H. Cottet, O. Theodoly, Macromolecules 40 (2007) 2672. DOI: 10.1021/ma062600+.
[14] C. Bauer, Rhodia internal technical report TR03-0088 and TR04-0005.
[15] S. Provencher, Comput. Phys. Commun. 27 (1982) 229.
[16] J.K. Percus, G. Yevick, Phys. Rev. 110 (1958) 1.
[17] M. Daoud, J.P. Cotton, J. Phys (Paris) 43 (1982) 531.
[18] J.S. Pedersen, M.C. Gerstenberg, Macromolecules 20 (1987) 1363.
[19] F. Muller, L. Delsanti, L. Auvray, P. Guenoun, J. Yang, Y.J. Chen, J.W. Mays, B. Démé, M. Tirrell, P. Guenoun, The European Physical Journal E **3** **(**2000) 45**.**
[20] P. Guenoun, F. Muller, M. Delsanti, L. Auvray, Y.J. Chen, J.W. Mays, M. Tirell, Phys. Rev. Lett. 81 (1998) 3872.
[21] J. Plĕstil, J. Appl. Cryst. 33 (2000) 600.
[22] G. Riess, Prog. Polym. Sci. 28 (2003) 1107.
[23] J. Grandjean, A. Mourchid, Physical Review E 72 (2005) Art. No. 041503.
[24] J. Loiseau, N. Doërr, J.M. Suau, J.B. Egraz, M.F. Llauro, C. Ladavière, J. Claverie, Macromolecules 36 (2003) 3066.
[25] Y. Dong, D.C. Sundberg, Journal of Colloid and Interface Science 258 (2003) 97.
[26] D. Bendejacq, Thesis of « université de Paris VI » (2002).
[27] D. Bendejacq, V. Ponsinet, M. Joanicot, Langmuir 21 (2005) 1712.
[28] M.W. Matsen, F.S. Bates, Macromolecules 29 (1996) 1091.
[29] A. Jada, G. Hurtrez, B. Siffert, G. Reiss, Macromol. Chem. Phys. 197 (1996) 3697.
[30] E.B. Zhulina, O.V. Borisov, Macromolecules 35 (2002) 9191.
[31] N. Dan, M.Tirell, Macromolecules 26 (1993) 4310.
[32] E.B. Zhulina, T.M. Birshtein, O.V. Borisov, Eur. Phys. J. E 20 (2006) 243.
[33] N. Gaillard, A. Guyot, J. Claverie, Journal of Polymer Science: Part A: Polymer Chemistry 41 (2003) 684.
[34] H. Matsuoka, S. Maeda, P. Kaewsaiha, K. Matsumoto, Langmuir 20 (2004) 7412.
[35] P. Kaewsaiha, K. Matsumoto, H. Matsuoka, Langmuir 21 (2005) 9938.
[36] I. Astafieva, X. Fu Zhong, A. Eisenberg, Macromolecules 26 (1993) 7339.
[37] S. Förster, M. Zisenis, E. Wentz, M. Antonietti, J. Chem. Phys 104 (1996) 9956.
[38] M. Jacquin, P. Muller, H. Cottet, R. Crooks, O. Théodoly, Controlling the Melting of Kinetically Frozen PBA-b-PAA Micelles via addition of surfactant, to appear in Langmuir.
[39] A. Morel, H. Cottet, M. In, S. Deroo, M. Destarac, Macromolecules 38 (2005) 6620.
[40] J.R.C. van der Maarel, W. Groenewegen, S.U. Egelhaaf, A. Laap, Langmuir 16 (2000) 7510.
[41] L. Willner, A. Poppe, J. Allgaier, M. Monkenbusch, D. Richter, Europhys. Lett. 55 (2001) 667.
[42] D. Bendejacq, M. Joanicot, V. Ponsinet, Eur. Phys. J. E17 (2005) 83.
[43] J. Israelachvili, Intermolecular &Surface Forces, 2nd ed. Academic Press (1992) chapter 17, p. 352.
[44] C. Marques, J.-F. Joanny, L. Leibler, Macromolecules 21 (1988) 1051.

**LIST OF FIGURES**

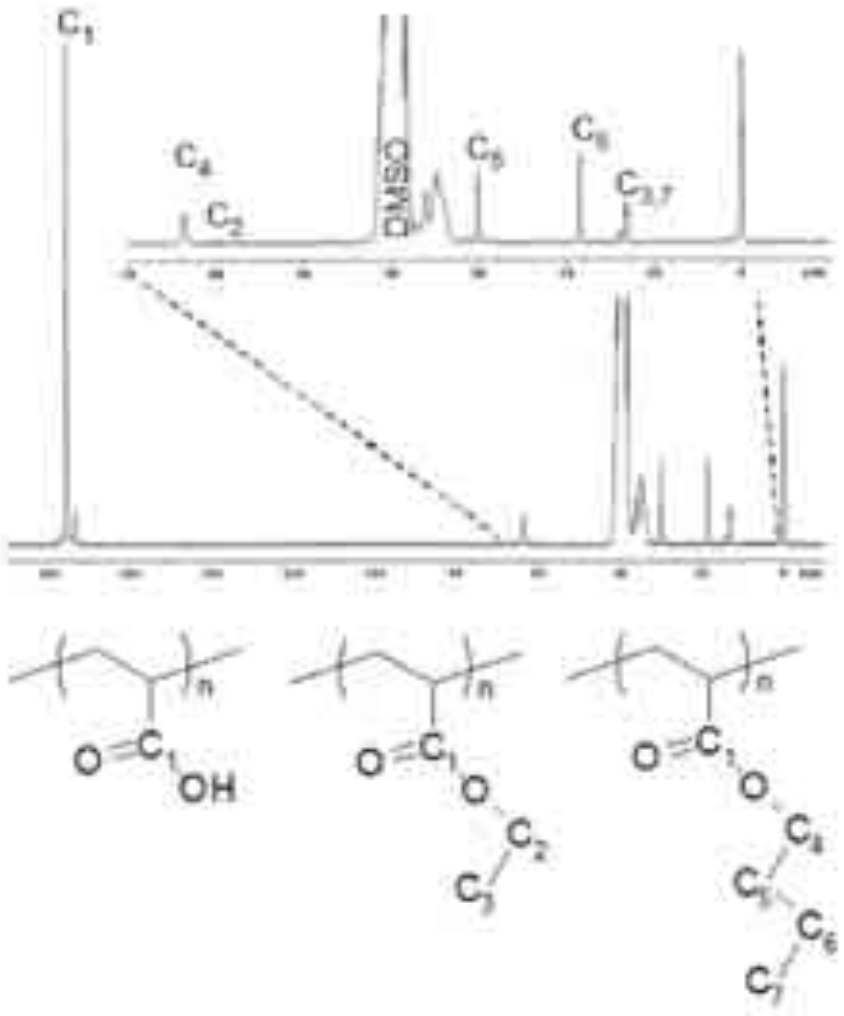

**Figure 1: $^{13}$C NMR spectrum of sample PBA-b-PAA 4k-20k in DMSO with Chromium acetyl acetonate with the peak indexed by their corresponding carbon on the developed chemical formula of the monomers.**

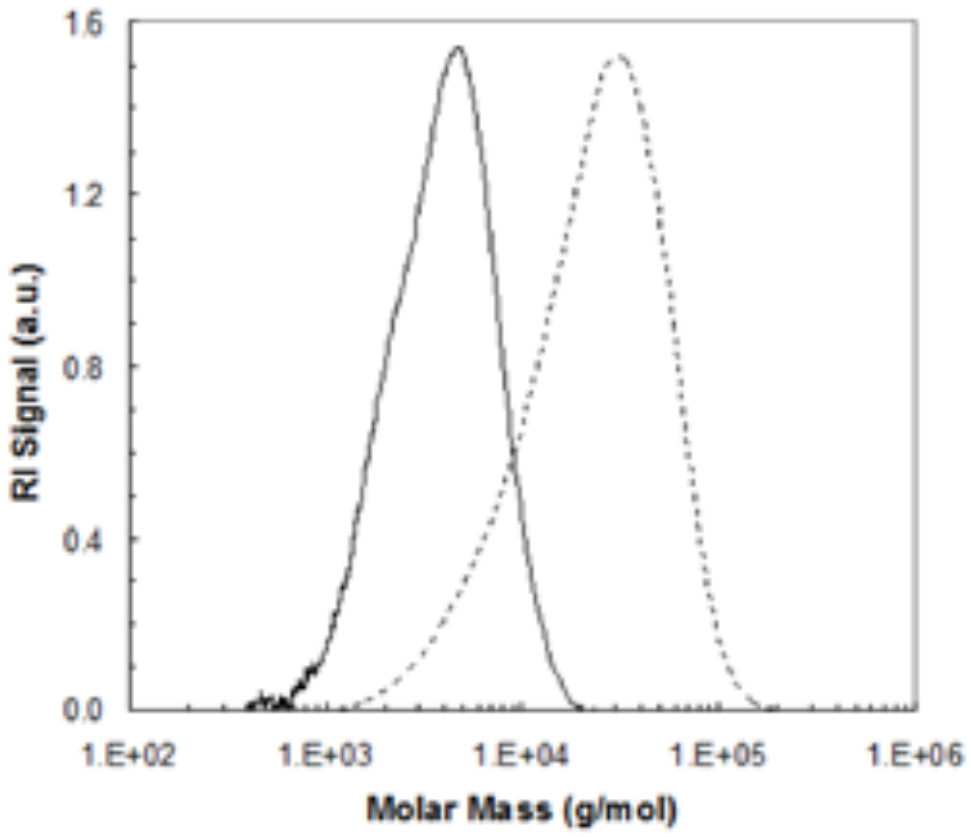

**Figure 2: GPC results of the aliquot of the first block PBA and the diblock after methylation for a PBA-b-PAA 4.5k-20k.**

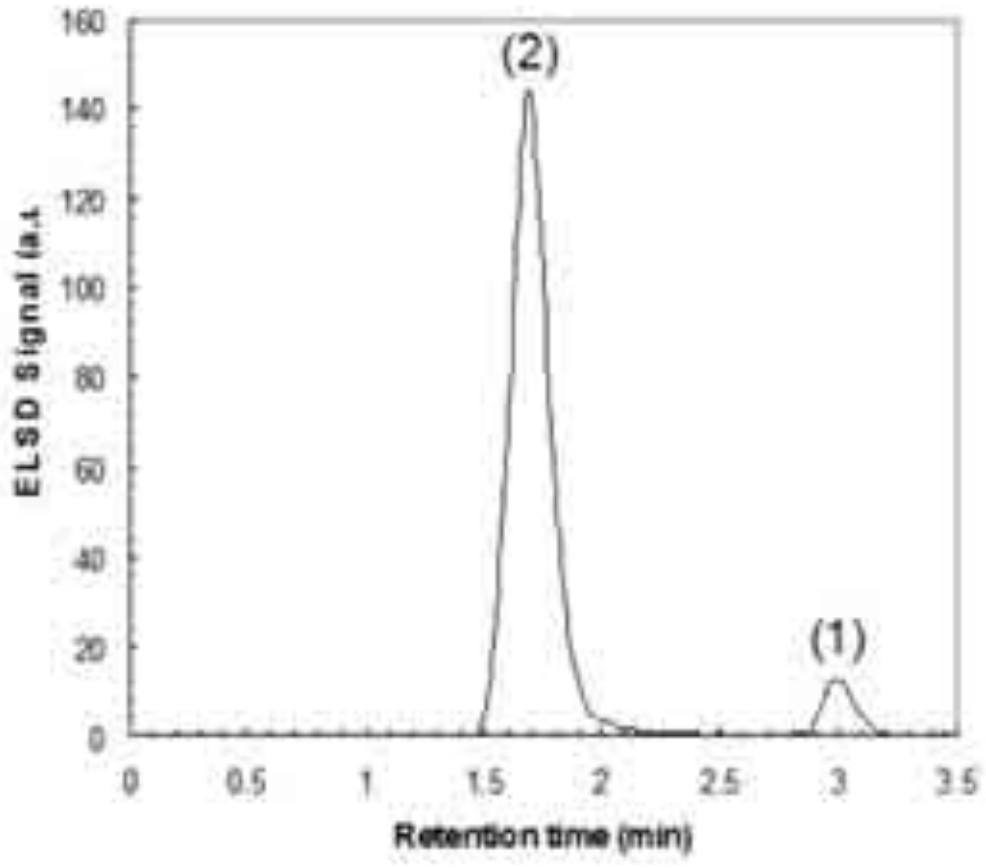

**Figure 3: LC-PEAT analysis of a PBA-b-PAA 6k-24k sample in THF/water 90/10 (v/v). Peak identification: peak (1), h-PBA; peak (2), PBA-b-PAA diblocks.**

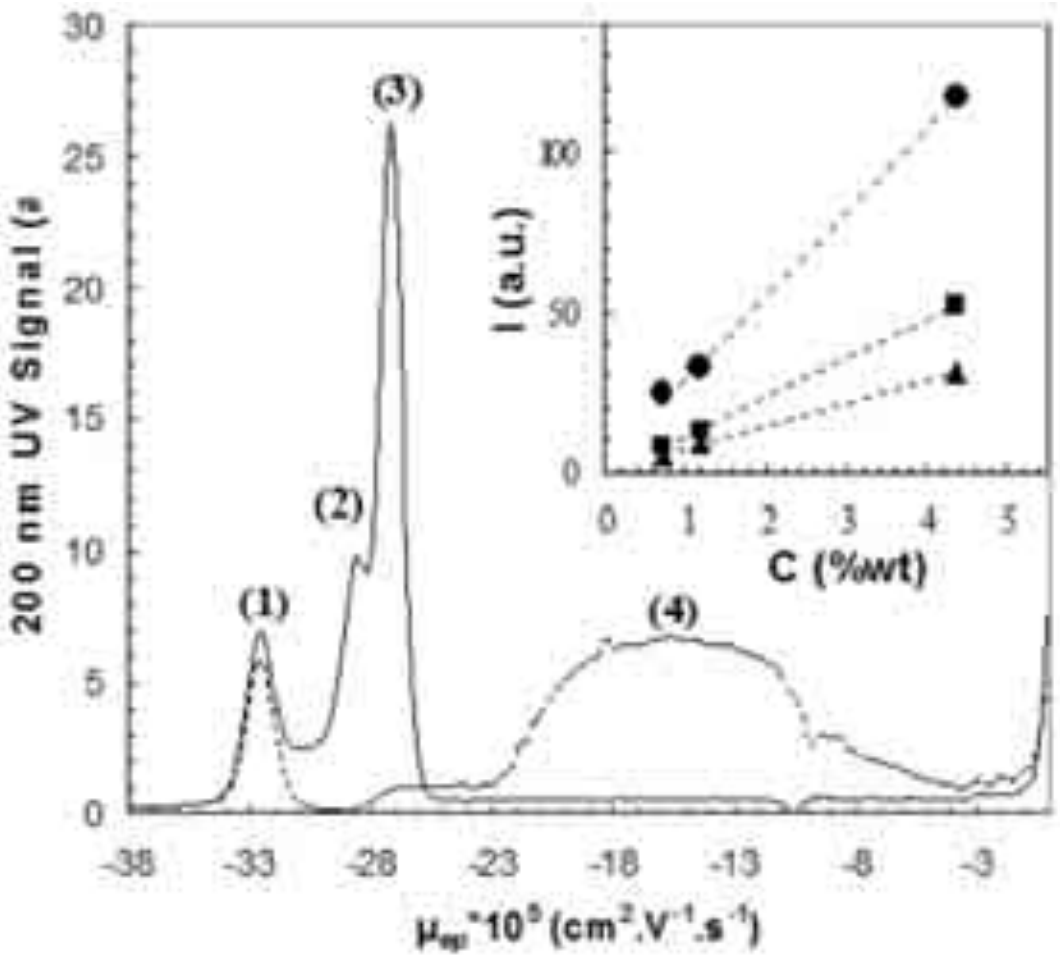

**Figure 4: CE Electropherogram of diblock PBA-b-PAA 3k-4k in a 160mM borate buffer as followed by UV detection at 200 nm. (—) without added surfactant and (- - -) with5 mM C12E6 surfactant in the electrolyte. Peak identification: peak (1), h-PAA; peak (2), PBA-b-PAA unimers; peak (3), PBA-b-PAA micelles, and peak (4) PBA-b-PAA diblocks associated with C12E6 surfactant. Insert: intensities of (●) peak (1), (■) peak (2), and (▲ ) peak (3) versus injected concentration.**

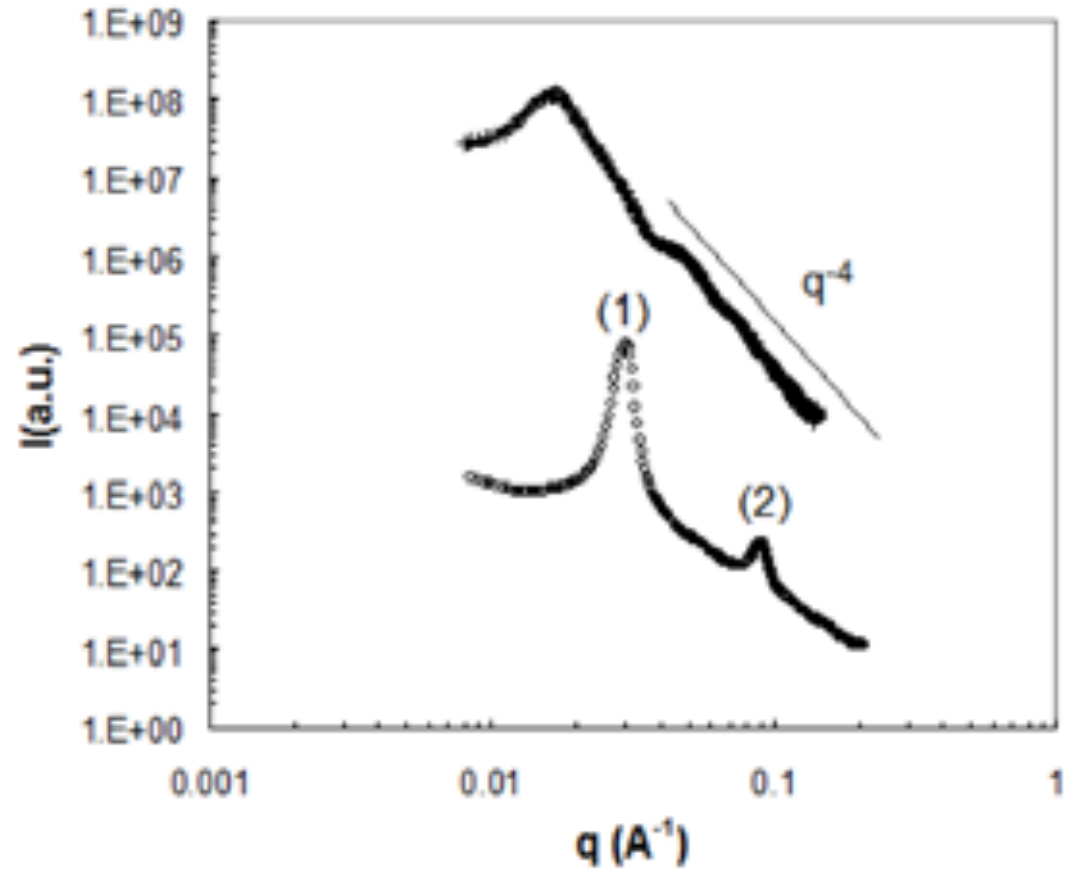

**Figure 5: SAXS scattering data for melt samples of PBA-b-PAA 6k-24k (+) and 3k-4k (o).**

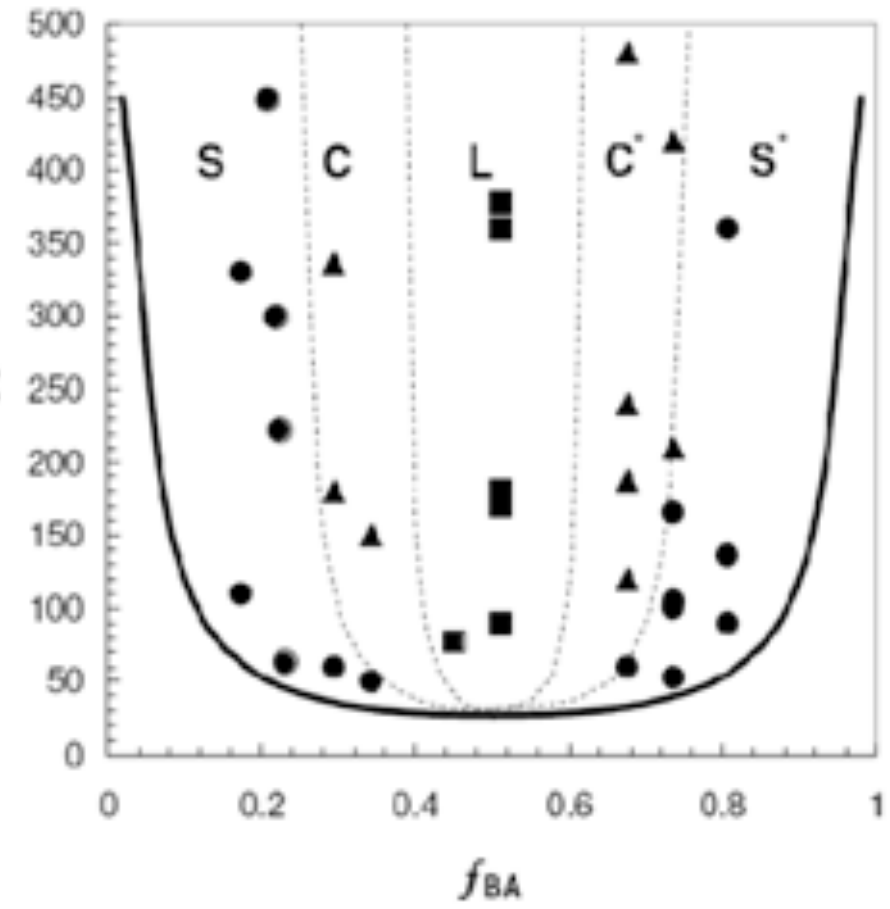

**Figure 6: Melt state structure topologies of diblock copolymers PBA-b-PAA versus the volume fraction $f_{BA}$ of BA in the diblock and the total length N of diblock chains in number of monomers. S stands for spheres, C for cylinders, L for lamellas, C* for inverted cylinders and S* for inverted spheres.**

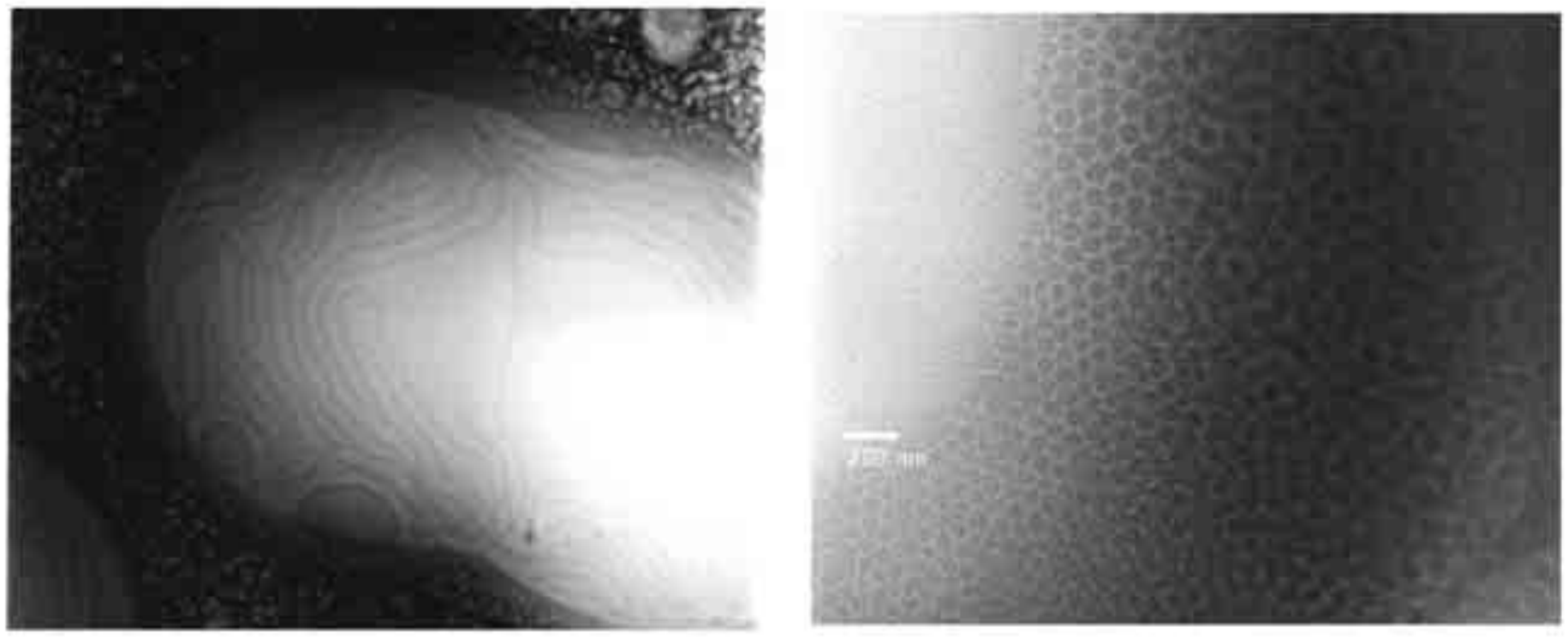

**Figure 7: Cryo-Tem pictures of PBA-b-PAA 3k-4k solutions in water at 2 wt% prepared from a) the cast film route b) the dialysis route.**

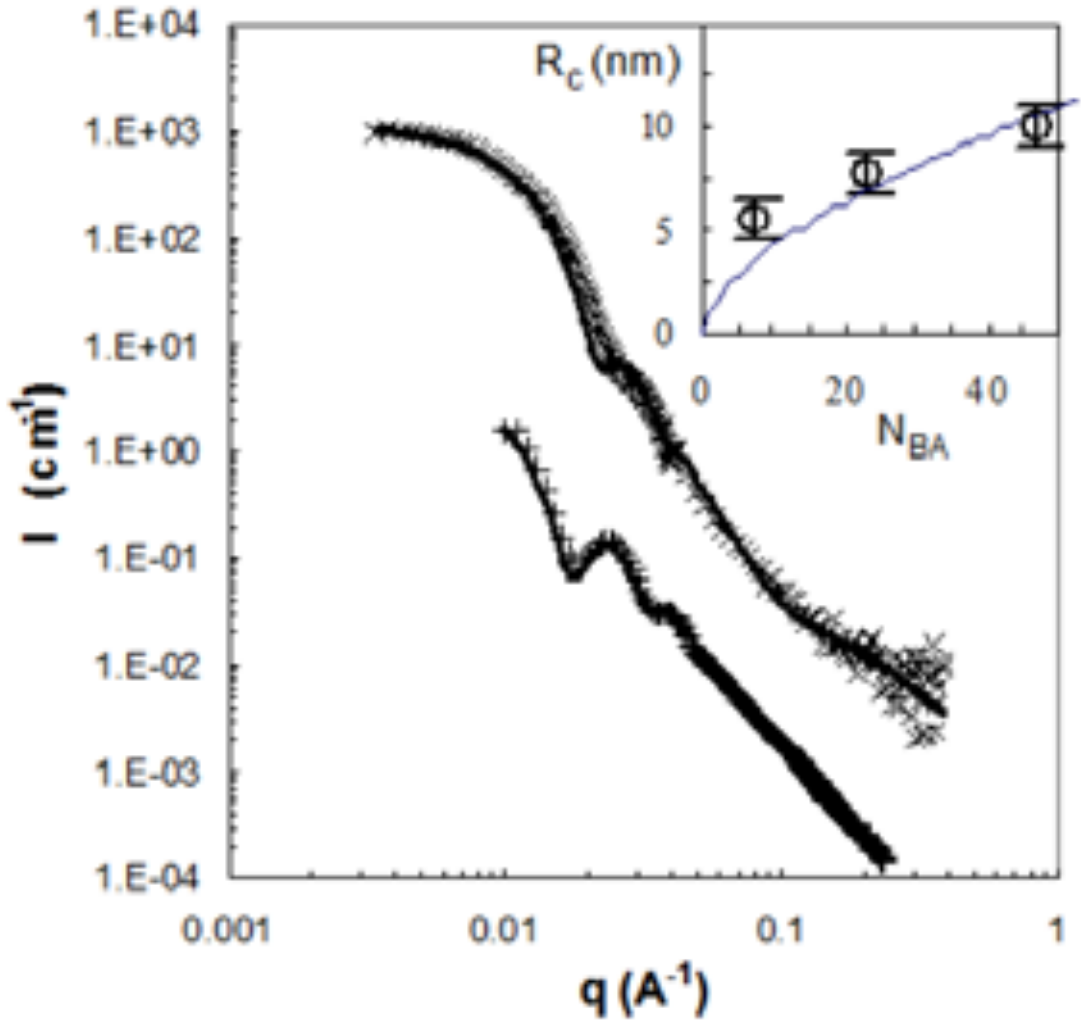

**Figure 8: SANS (×) and SAXS (+) scattering data of PBA-b-PAA 3k-4k solutions in $D_2O$ at 2 wt%. Insert: Core radii $R_c$ versus PBA degree of polymerisation $N_{BA}$ for PBA-b-PAA diblocks with $X_{BA/AA}$=0.25.**

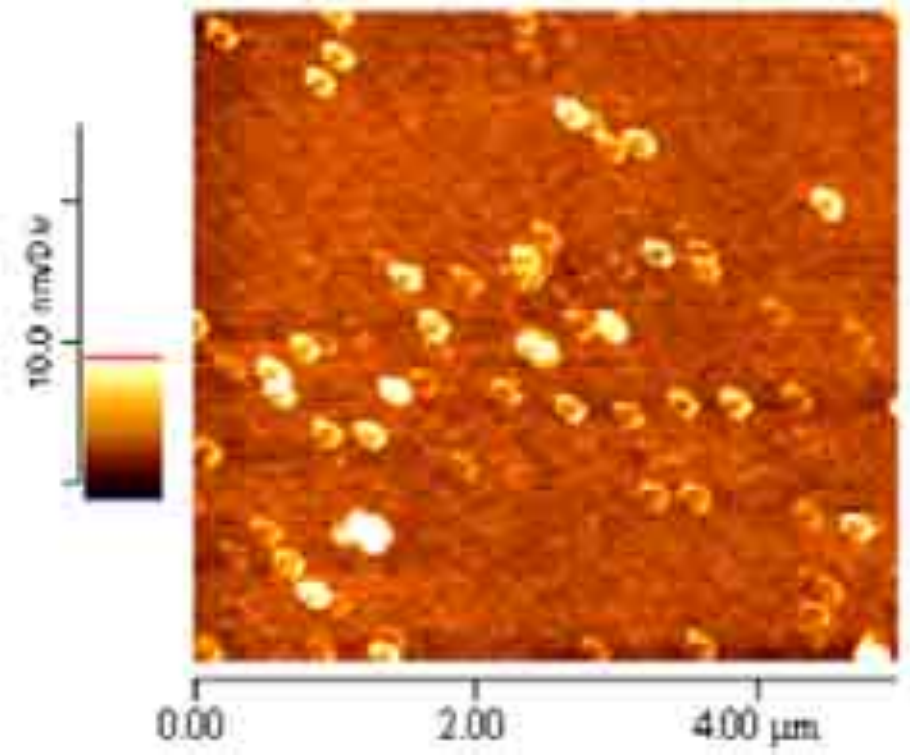

**Figure 9: AFM picture in water of PBA-b-PAA micelles adsorbed on a cationic surface in solution at** $\alpha$=1.

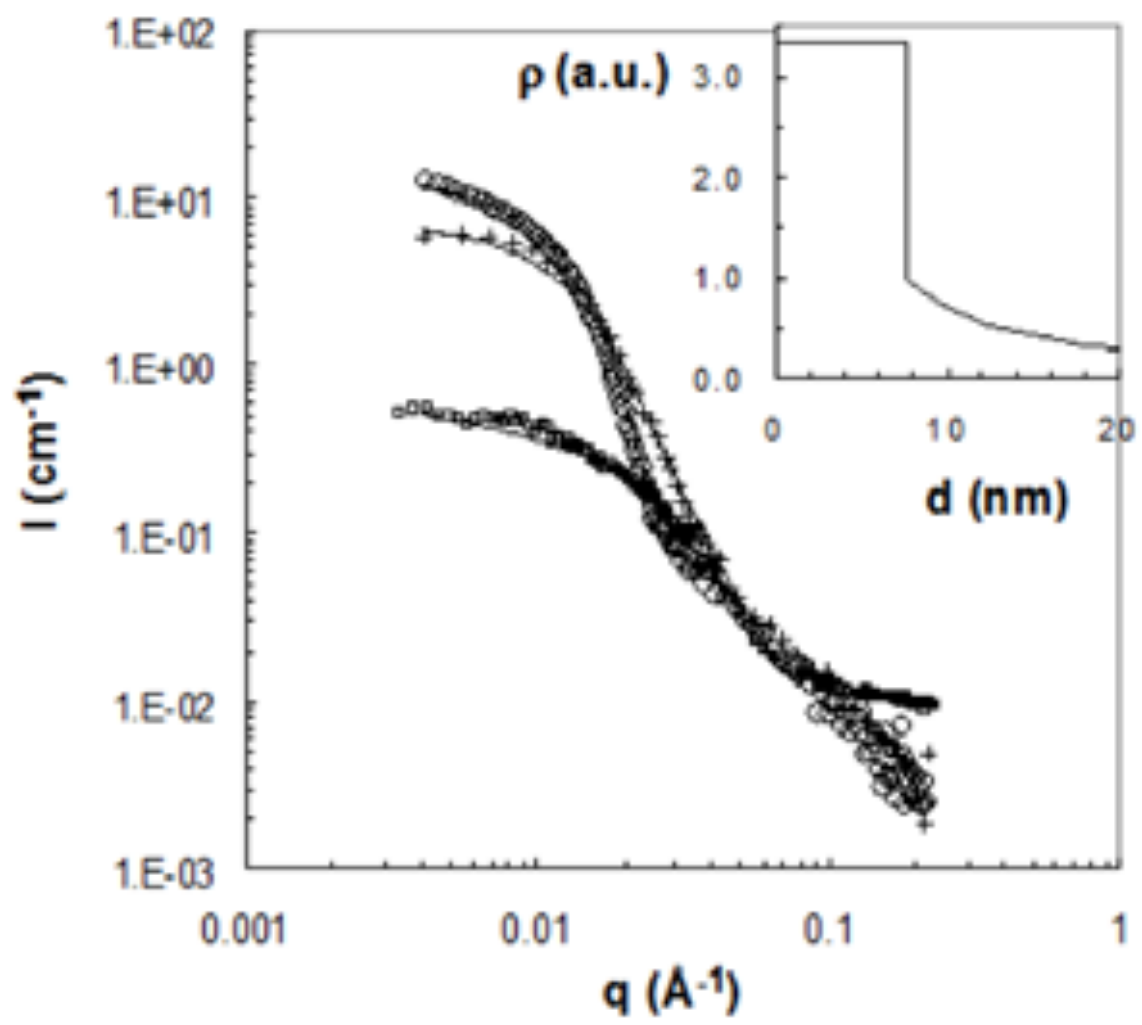

**Figure 10: SANS data of $d_9$PBA-b-PAA 3k-12k (♣) at pH=8 and C=1.5 wt% in $D_2O$, (❡) at $\alpha$=0 in $D_2O$, and (+) at $\alpha$=0 in D2O/ H2O 82/18 (v/v). Solid lines: adjustment of data by Daoud cotton core-corona model at $\alpha$=0 and Urchin model at $\alpha$=1. Insert: Contrast profile used for adjustment of data at $\alpha$=0.**

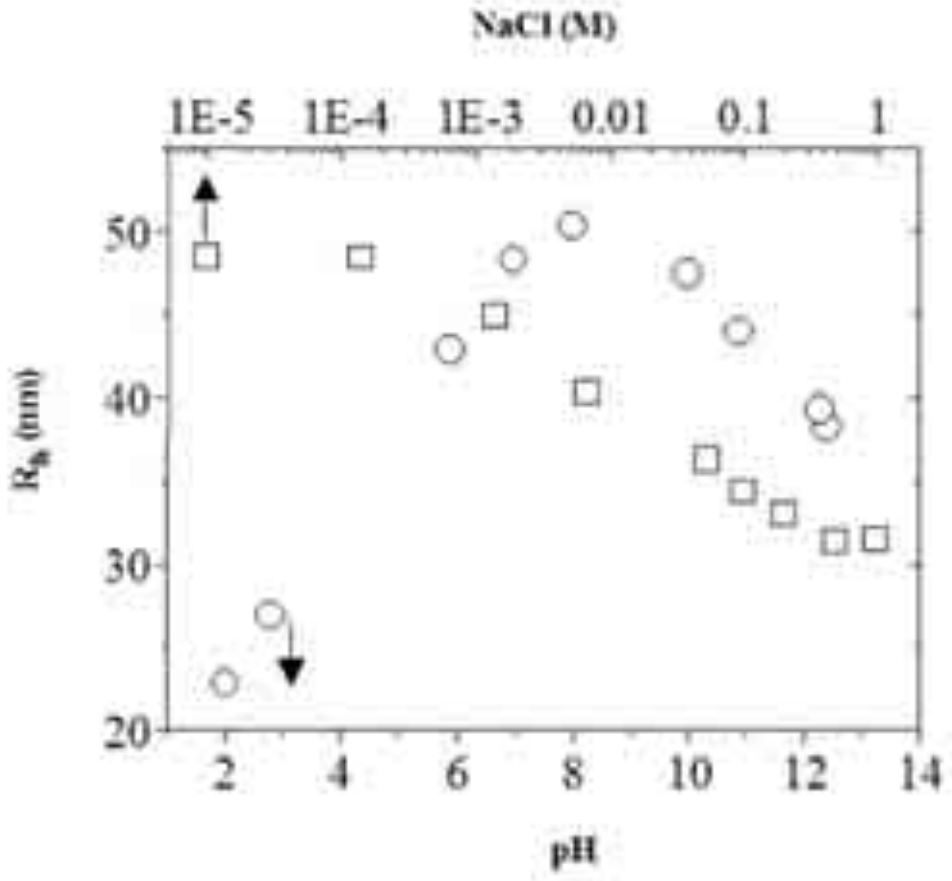

**Figure 11: Hydrodynamic radii measured for PBA-b-PAA 3k-12k by DLS (❡) versus pH without added NaCl, and (♣) versus NaCl concentration at $\alpha$=1.**

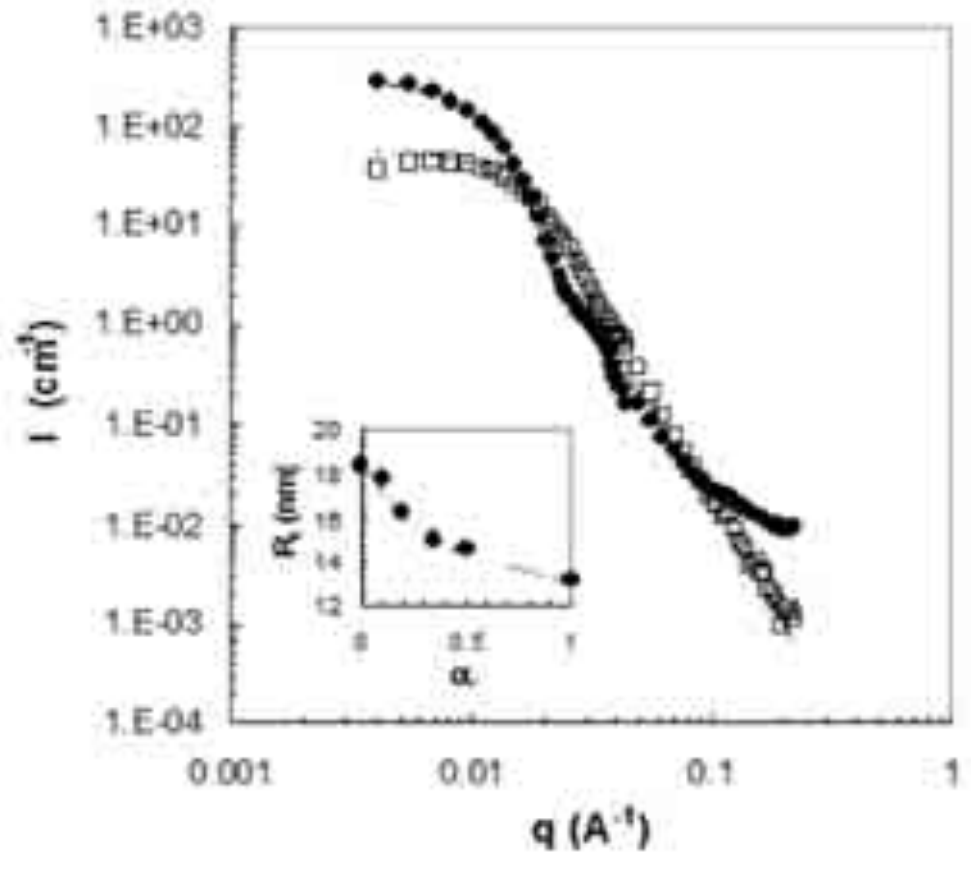

**Figure 12: SANS data of PBA-b-PAA 3k-4k at C=0.5 wt% and ionization (❡) α=0, and (●) α=1. Solid lines correspond to adjustments by Pedersen model for α=0 and polydisperse spheres for α=1. Insert: Evolution of the core radius versus ionization for PBA-b-PAA 3k-4k.**

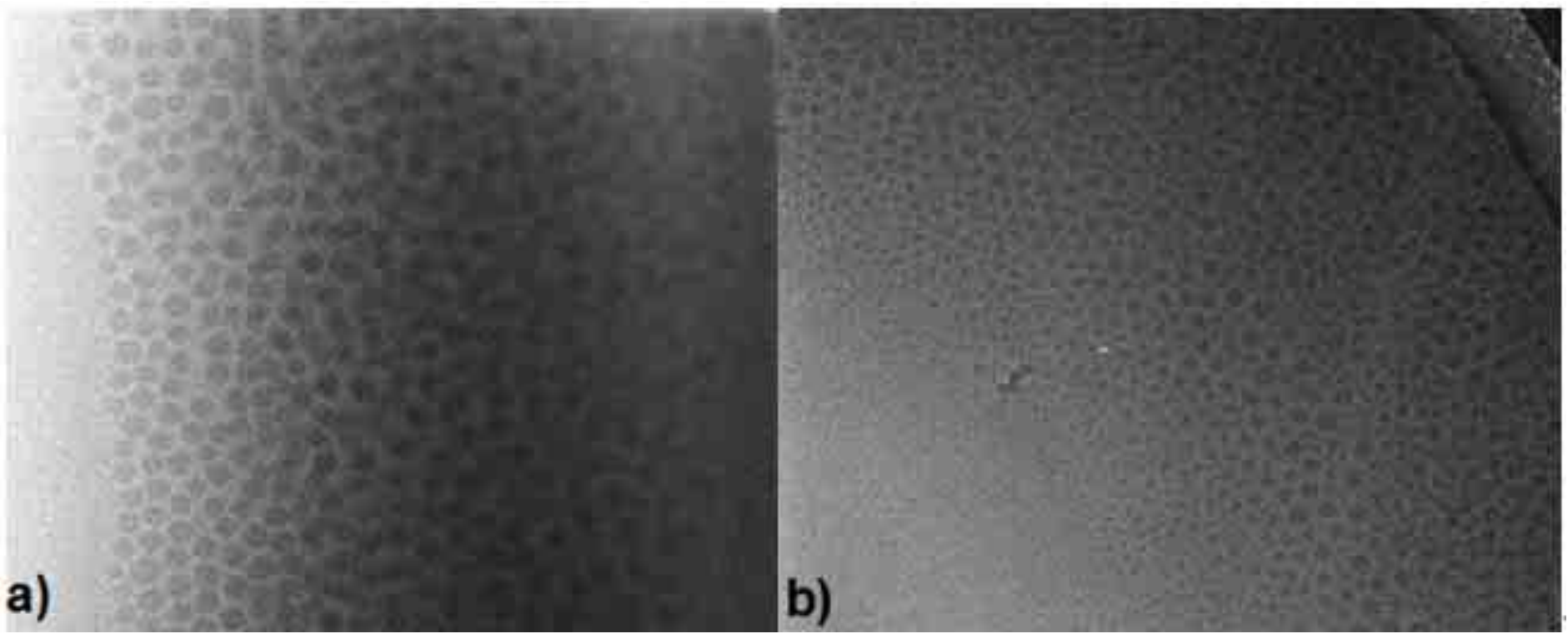

**Figure 13: Cryo-TEM pictures of PBA-b-PAA 3k-4k at C=2 wt% and ionization a) α=0, and b) α=1.**

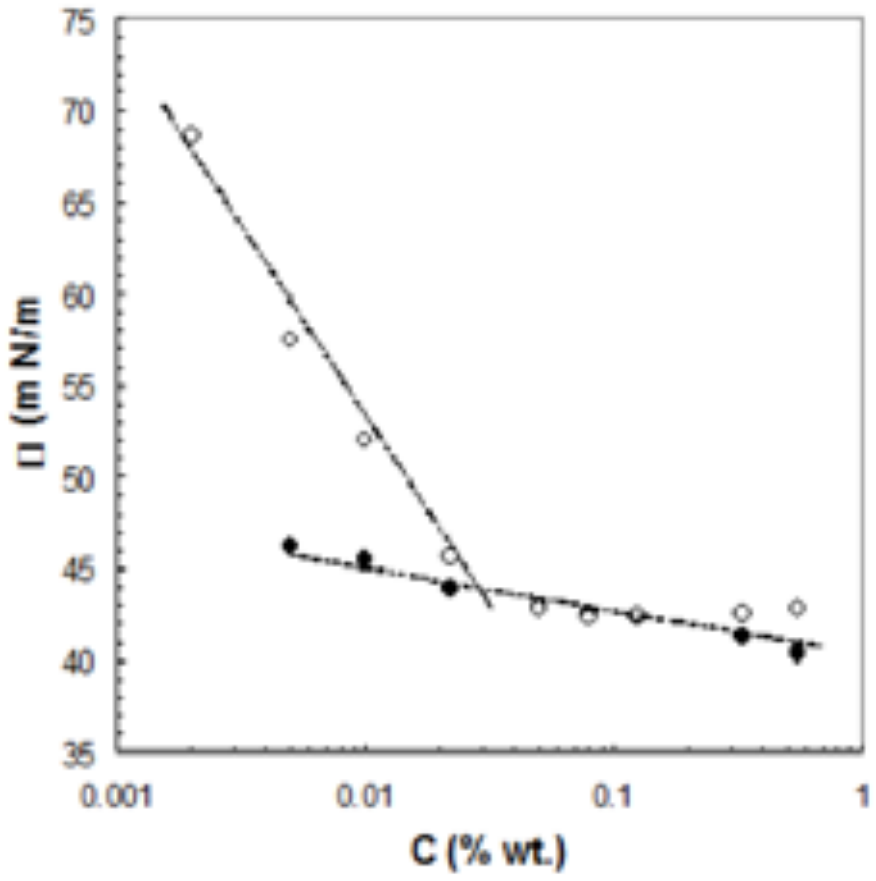

**Figure 14: Surface tension of PBA-b-PAA 1k-4k solution at α=1 versus concentration measured (❡) one hour after creation of a fresh interface, and (●) at equilibrium.**

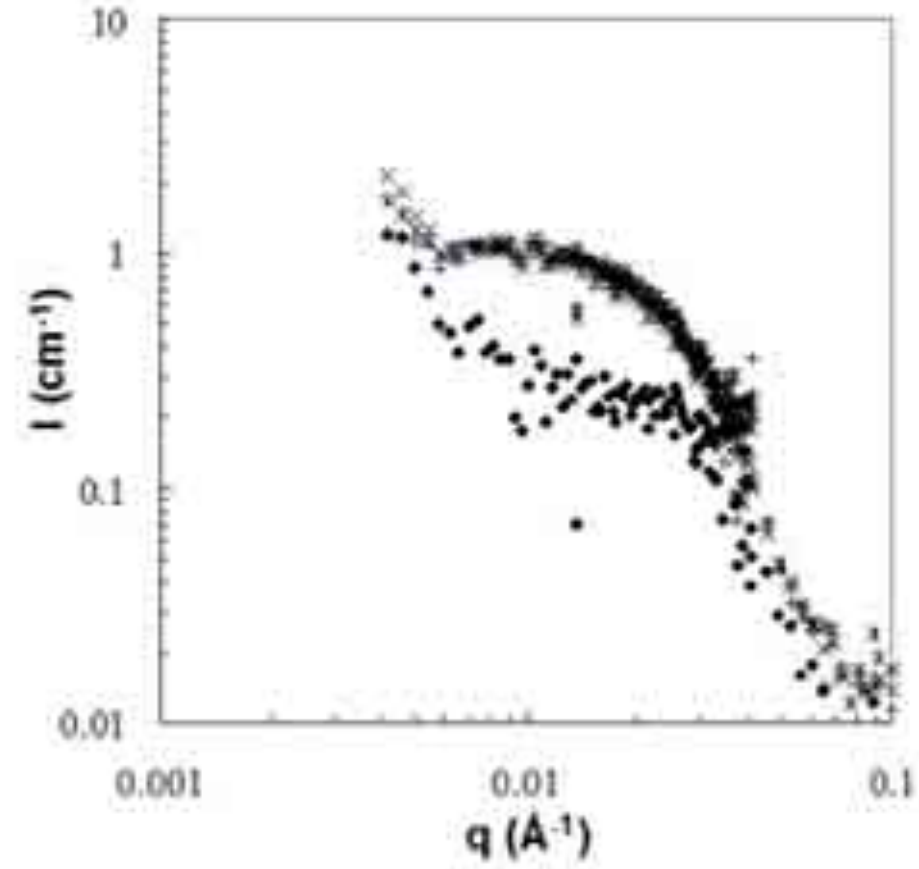

**Figure 15: SANS data of mixtures of PBA-b-PAA 3k-12k and $d_9$PBA-b-PAA 3k-12k (+) solution t = 0, (×) solution t = 3 days, (✱) solution t = 6 months, and (●) solution t=∞.**

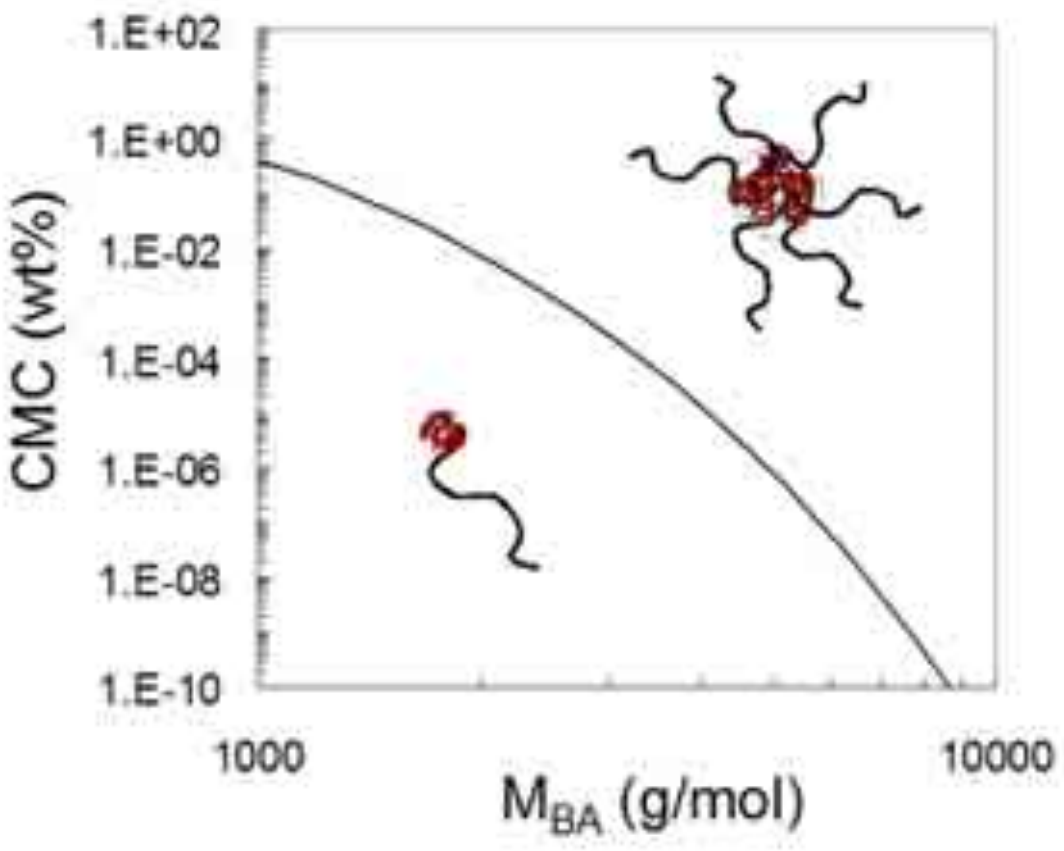

**Figure 16: Calculated CMC values for PBA-b-PAA diblocks versus the molar mass of the PBA block.**